\documentclass[english,11pt,oneside]{article}
\pdfoutput=1
\usepackage{amsmath}
\usepackage{amsfonts}
\usepackage{amssymb}
\usepackage{graphicx, rotating}
\usepackage{epstopdf}
\usepackage{epsfig}
\usepackage{latexsym}
\usepackage{multirow}
\usepackage{color}
\usepackage{slashed,cite}
\usepackage[top=2.8 cm, bottom=2.8 cm, left=3 cm, right=3 cm]{geometry}
\usepackage{times}

\newcommand{\ba}{\begin{eqnarray}}
\newcommand{\ea}{\end{eqnarray}}
\newcommand{\no}{\nonumber}
\usepackage[colorlinks]{hyperref}
\hypersetup{
    citecolor={blue}
}

\begin{document}
\title{Solar Neutrinos as a Probe of Dark Matter-Neutrino Interactions}

\author{Francesco Capozzi$^{a,b,c}$\footnote{capozzi.12@osu.edu}~~~~~~~Ian M. Shoemaker$^{d}$\footnote{ian.shoemaker@usd.edu}~~~~~~~~Luca Vecchi$^{a,b}$\footnote{vecchi@infn.pd.it}\\
{\small\emph{$^a$ Dipartimento di Fisica e Astronomia, Universit\`a di Padova and}}\\
{\small\emph{$^b$INFN, Sezione di Padova, via Marzolo 8, I-35131 Padova, Italy}}\\
{\small\emph{$^c$Center for Cosmology and AstroParticle Physics (CCAPP), Ohio State University, Columbus, OH 43210}}\\
{\small\emph{$^d$Department of Physics, University of South Dakota, Vermillion, SD 57069, USA}}
}

\date{}
\maketitle

\begin{abstract}
Sterile neutrinos at the eV scale have long been studied in the context of anomalies in short baseline neutrino experiments. Their cosmology can be made compatible with our understanding of the early Universe provided the sterile neutrino sector enjoys a nontrivial dynamics with exotic interactions, possibly providing a link to the Dark Matter (DM) puzzle. Interactions between DM and neutrinos have also been proposed to address the long-standing ``missing satellites'' problem in the field of large scale structure formation. Motivated by these considerations, in this paper we discuss realistic scenarios with light steriles coupled to DM. We point out that within this framework active neutrinos acquire an effective coupling to DM that manifests itself as a new matter potential in the propagation within a medium of asymmetric DM. Assuming that at least a small fraction of asymmetric DM has been captured by the Sun, we show that a sizable region of the parameter space of these scenarios can be probed by solar neutrino experiments, especially in the regime of small couplings and light mediators where all other probes become inefficient. In the latter regime these scenarios behave as familiar $3+1$ models in all channels except for solar data, where a Solar Dark MSW effect takes place. Solar Dark MSW is characterized by modifications of the most energetic $^8$B and CNO neutrinos, whereas the other fluxes remain largely unaffected. 

 \end{abstract}

\newpage

\section{Motivation}
\label{sec:Motiv}

Dark Matter and neutrino oscillations represent unquestionable evidence of physics beyond the Standard Model. The missing matter problem is solved postulating the existence of a yet to be discovered exotic, dark sector containing long-lived particles. A very attractive framework is offered by WIMP (weakly-interacting massive particle) dark matter, within which the observed abundance as well as the gross features of the dark matter distribution can be accommodated if the WIMP annihilation cross section into visible particles is of the order of a picobarn.

The most convincing explanation of neutrino oscillations is that neutrinos have non-vanishing masses. The fundamental origin of neutrino masses requires new particles or new mass thresholds, and a precise investigation of neutrino oscillations may provide important hints on such new physics. Interestingly, a number of experimental anomalies in short baseline experiments has been observed, pointing to the existence of a new sterile neutrino with mass in the eV range~\cite{Gariazzo:2015rra,Kopp:2013vaa} and a sizable mixing with standard active neutrinos. While awaiting for a definitive experimental confirmation, it is however important to theoretically investigate the phenomenological implications of the new fermions. A first important aspect is that such sterile neutrinos must be part of a more involved, secluded sector in order to be consistent with cosmology ~\cite{Babu:1991at,Hannestad:2013ana,Dasgupta:2013zpn,Mirizzi:2014ama,Cherry:2014xra,Chu:2015ipa,Cherry:2016jol,Vecchi:2016lty}. A tantalizing possibility is that sterile neutrinos may directly communicate with (or be themselves part of) the dark matter sector. Curiously, this conjecture fits well with a set of independent arguments. Indeed, it turns out that an apparent inconsistency between observations and simulations of large scale structure, the so-called ``missing satellite problem", may be reconciled with the usual WIMP paradigm by postulating the presence of additional interactions between dark matter (DM) and neutrinos, or light exotic neutrinos, that keeps the dark matter kinetically coupled until much later than usual cold candidates~\cite{Hooper:2007tu,Aarssen:2012fx}. Extensions and further implications of scenarios such as these have also been studied in \cite{Shoemaker:2013tda,Dasgupta:2013zpn,Cherry:2014xra,Bertoni:2014mva,Binder:2016pnr}.

\begin{figure*}[t]
\begin{center}
\includegraphics[width=7cm]{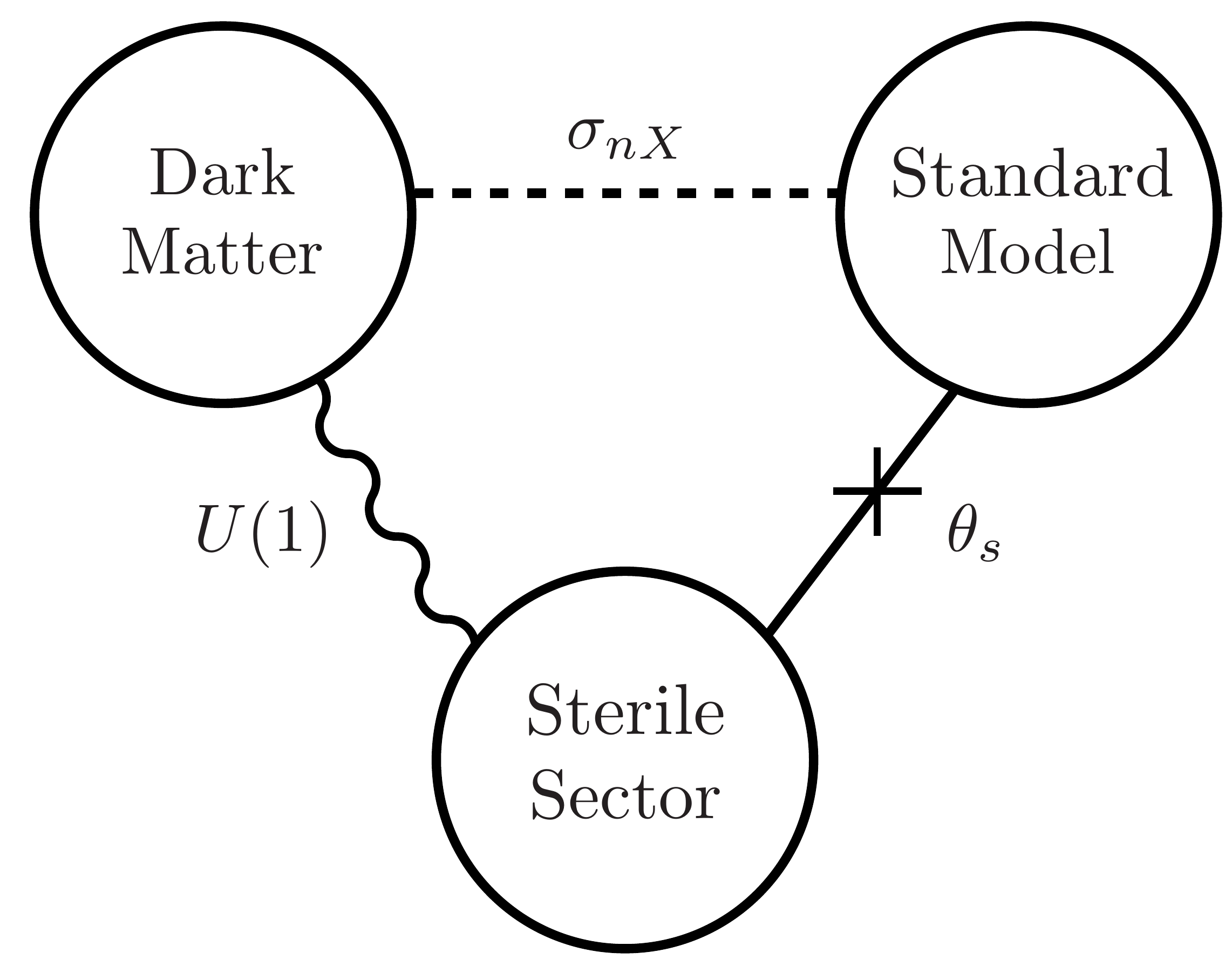}
\caption{Schematic picture of the framework considered in this paper. See Section~\ref{sec:model} for a detailed description. 
}
\label{fig:scheme}
\end{center}
\end{figure*}

Upon putting these pieces together, a consistent picture seems to emerge, see Fig.~\ref{fig:scheme}. WIMP dark matter must interact with the Standard Model in order to possess the observed abundance. The same interactions responsible to generate an efficient annihilation into visible particles are also expected to show up as a non-vanishing rate for scattering off nucleons, $\sigma_{nX}$, that may represent our best hope to observe DM directly. Our world further sees a Sterile Sector via a mixing angle $\theta_s$ between active neutrinos and steriles at the eV scale. The sterile neutrinos are part of a more involved sector that communicates with the DM, thus playing the role of an additional, novel link between visible and dark matter. 

While the various links depicted in Fig.~\ref{fig:scheme} have been already independently investigated, no work has ever considered the entire picture and identified its distinctive signatures.This is the aim of our work. We find that if the DM is a standard {\emph{symmetric}} species, the scheme of Fig.~\ref{fig:scheme} reveals no qualitatively new phenomenon and is best probed by an investigation of the individual links. On the other hand, in scenarios in which the abundance of DM particles and anti-particles is different, the new DM coupling to the Sterile Sector may be observed as a novel matter effect in the propagation of active neutrinos. We find that in the limit of light mediators this ``{\it Dark MSW}'' phenomenon becomes the smoking gun of such DM-neutrino interactions, and therefore of the scheme of Fig.~\ref{fig:scheme}. In this paper we investigate the Dark MSW effect and therefore decide to focus on scenarios of asymmetric WIMP Dark Matter (see e.g.~\cite{Kaplan:2009ag} and references therein).


DM effects in the propagation of high energy neutrinos through the DM halo have been previously discussed in~\cite{Horvat:1998ym,Lunardini:2000swa,Miranda:2013wla,deSalas:2016svi}. However, our poor knowledge of the flavor content of these neutrinos does not allow us to place robust constraints. A more promising way to observe the Dark MSW effect is obtained if some DM particles have been captured by the Sun. In this case a Solar Dark MSW phenomenon can impact solar neutrinos, the composition of which is presently well known. Dark matter capture by the Sun is a generic expectation in the presence of a non-negligible coupling to nucleons, $\sigma_{nX}$, and has been considered extensively in the literature as potential sources of indirect signatures of dark matter as well as to alleviate the so-called ``solar abundance problem" (see e.g.~\cite{Cumberbatch:2010hh,Taoso:2010tg,Frandsen:2010yj,Vincent:2015gqa}). Here we show that, even if only a tiny fraction of asymmetric dark matter is present in the sun, solar data may represent an ideal probe of the DM-neutrino interactions in the scheme of Fig.~\ref{fig:scheme}.

In Section~\ref{sec:model} we introduce our framework and discuss the main astrophysics constraints. The impact on solar neutrinos is analyzed in detail in Section~\ref{sec:DarkMSW}. We conclude in Section~\ref{sec:conc}.

\section{The Model and Existing Constraints}
\label{sec:model}

Our framework is schematically depicted in Fig.~\ref{fig:scheme}. We postulate the existence of two exotic sectors beyond the Standard Model: an Asymmetric Dark Matter and a Sterile Neutrino Sector.

We assume the dark matter (DM) is a fermionic WIMP and denote it by $X$. No qualitative change would arise with a bosonic candidate. More precisely, as anticipated in the Introduction, we
consider scenarios in which the DM abundance is set by an asymmetry as in ``asymmetric DM'' models~\cite{Kaplan:2009ag}. As all WIMPs, our candidate has non-negligible couplings to ordinary matter, that are necessary to guarantee thermal equilibrium in the early Universe and induce a relic abundance via the familiar thermal freeze-out paradigm. In addition, asymmetric DM scenarios have two characterizing features~\cite{Kaplan:2009ag}. First, the typical DM mass lies in the $5-15$ GeV range. Second, in addition to the interactions mediating freeze-out, these models must possess number-changing operators that transfer the asymmetry between visible and dark sectors. It follows that a general, model-independent expectation is a non-negligible rate for scattering off nucleons $\sigma_{nX}$. The explicit form of the short distance interactions inducing the latter is model-dependent and not relevant here. For definiteness, $\sigma_{nX}$ may be assumed to be mediated by higher-dimensional operators.


The Sterile Neutrino Sector contains one (or more) sterile neutrinos $\nu_s$ that mix with the active neutrinos. Motivated by anomalies in short-baseline neutrino experiments, we will consider sterile masses $m_4\sim$ eV and mixing angles with active species of order $\sin\theta_s\sim0.1$. Our main results of Section~\ref{sec:DarkMSW} do however extend way outside this domain (see also Section~\ref{sec:conc}).

As anticipated in the Introduction, sterile neutrinos compatible with Cosmology must be part of a non-trivial sector with exotic interactions. We model the latter introducing a gauge $U(1)$ carried by $\nu_s$ and the Dark Sector, parametrized for simplicity by a single DM field $X$. To allow a mixing between $\nu_s$ and active neutrinos, compatibly with the gauge symmetries, we further introduce right-handed fermions $N$ that are singlets under all gauge symmetries as well as a sterile scalar $\phi$ with $U(1)$ charge conjugate to $\nu_s$, so that $\phi\nu_s$ can form a complete singlet (analogously to the Standard Model $HL$) and couple to $N$ via a renormalizable interaction. In formulas, neglecting the kinetic terms and suppressing gauge, family, and Lorentz indices for brevity, the new interactions read:~\footnote{For definiteness we assume that $\nu_s$ is a Dirac fermion with a chiral symmetry that forbids a mass term. The reader who does not like global symmetries is invited to have a look at Appendix \ref{sec:anomalyfree}, where a model in which the chiral symmetry appears accidentally is presented. The point of view promoted here is however that (\ref{Lagrangian}) should be intended as a toy model describing a larger class of scenarios with sterile-DM interactions. The results of our paper are largely independent on the details of the model: the key ingredients are those pictorially described in fig. \ref{fig:scheme}.}
\ba\label{Lagrangian}
{\cal L}&\supset&g_AA'_\mu J^\mu\\\no
&+&y_s\overline{N}\phi\nu_s+y_a\overline{N}HL+{\rm hc}\\\no
&+&\frac{m_N}{2}\overline{N}N^c+{\rm hc}.
\ea
Here $H$ is the SM Higgs, $L$ the lepton doublet, and
\ba
J^\mu=q_s\overline{\nu_s}\gamma^\mu\nu_s+q_X\overline{X}\gamma^\mu X+J_\phi^\mu,
\ea
with $q_s,q_X$ the corresponding sterile and DM charges under the exotic $U(1)$ and $J_\phi^\mu$ the current associated to $\phi$. We neglect a possible kinetic mixing with hypercharge as well as $\lambda_{\phi h}|\phi|^2|H|^2$ because their presence would not affect our conclusions. The main role of $\phi$ is to spontaneously break the $U(1)$. Fluctuations of the scalar around its vacuum expectation value are not relevant to the analysis of Section \ref{sec:DarkMSW}, that is the main subject of the paper.

The vacuum expectation value of $\phi$ generates a mass $m_A$ for $A'_\mu$ and induces a $N,\nu_s$ mixing. Similarly, after electroweak symmetry breaking the second line of Eq.~(\ref{Lagrangian}) in turn induces a mixing between active neutrinos and $N$. With $m_N\gg y_a\langle H\rangle, y_s\langle\phi\rangle$ the neutrinos are Majorana, whereas they are pseudo-Dirac when $m_N$ is small. Whether neutrinos are Dirac or Majorana will not affect our conclusions. The important point is that $\nu_s$ mixes with the active neutrinos $\nu_a\subset L$, thus transfering to the actives an exotic interaction with the $U(1)$ gauge field. Barring unnatural cancellations, the mixing angle is of order $\sin^2\theta\sim{\rm min}({y_a\langle H\rangle/y_s\langle\phi\rangle},{y_s\langle\phi\rangle/y_a\langle H\rangle})$. For our benchmark scenario $y_s\langle\phi\rangle> y_a\langle H\rangle$, so the heavy mass eigenstate is mostly sterile.

For clarity we emphasize that Eq.~(\ref{Lagrangian}) only captures the interactions relevant to the Dark Sector-Sterile Sector link. The dominant interactions between the DM and the SM (represented by $\sigma_{nX}$ in Fig.~\ref{fig:scheme}) are highly model-dependent and were not specified in Eq.~(\ref{Lagrangian}).

\subsection{Constraints}
\label{sec:constr}

In this section we discuss the constraints on the various links in Fig.~\ref{fig:scheme}. In evaluating the bounds we assume $m_4=1$ eV, $\sin^2\theta_s=0.01$, $|q_s|=|q_X|/2=1$ and $m_{X}=5$ GeV. We relax these assumptions in Section~\ref{sec:DarkMSW}. The constraints obtained in this section are reported in Fig. \ref{fig:dMSW} in terms of 
the gauge boson mass, $m_{A}$, and the gauge coupling, $g_{A}$.

The constrained region (filled in yellow) leaves a large portion of the parameter space unconstrained. However, as argued in Section~\ref{sec:DarkMSW}, Dark MSW takes place above the red curves. Therefore neutrino experiments have the potential to probe interesting regions of parameter space.

\subsubsection{The $\sigma_{nX}$ link}

Our DM candidate is asymmetric, which means that the present-day population is composed dominantly of particles and basically no antiparticles. Therefore indirect signatures of DM annihilation into neutrinos or other SM particles are suppressed (though potentially detectable if the present-day asymmetry is not too large~\cite{Murase:2016nwx}). In such a framework direct detection experiments, designed to probe the DM scattering rate off nucleons $\sigma_{nX}$, represent the most obvious probes of our Dark Sector. Current bounds for the mass range of interest, $m_X\sim5-15$ GeV, conservatively read $\sigma_{nX}\lesssim10^{-45}$ cm$^2$ for spin-independent rates \cite{xenon1T}. Spin-dependent interactions are far less constrained by direct detection experiments and allow cross sections up to $\sigma_{nX}\sim 10^{-38}~{\rm cm}^{2}$~\cite{Amole:2016pye}.

\subsubsection{The $\theta_s$ link}
\label{motivated}

The Sterile-Standard Model link is constrained by neutrino oscillation experiments. Throughout the paper we will adopt the region of the parameter space around ($\sin\theta_s\sim0.1, m_4\sim1$ eV), motivated by anomalies in short baseline experiments, as reported in recent global analyses ~\cite{Gariazzo:2015rra,Kopp:2013vaa}. Nevertheless, our results apply to a wider class of models.

Sterile neutrinos are constrained also by Cosmology. Indeed, for the values
of mass and mixing we are considering sterile neutrinos are fully thermalized via oscillations during the BBN and the CMB epoch, resulting in a tension with existing cosmological bounds on 
the radiation density and neutrino mass \cite{sterile_cosmology_1,sterile_cosmology_2,sterile_cosmology_3}. However, this conclusion strictly applies to ``truly sterile" neutrinos only. It is possible to inhibit the formation of a large population of sterile neutrinos if the latter have exotic interactions~\cite{Babu:1991at,Hannestad:2013ana,Dasgupta:2013zpn,Mirizzi:2014ama,Cherry:2014xra,
Chu:2015ipa,Cherry:2016jol,Vecchi:2016lty}. However, the underlying mechanism is strongly model-dependent and the associated bounds are as well. Instead of discussing the constraints for all the available models  (see~\cite{Babu:1991at,Hannestad:2013ana,Dasgupta:2013zpn,Mirizzi:2014ama,Cherry:2014xra,Chu:2015ipa,Cherry:2016jol,Vecchi:2016lty} for details) we find it more useful to simply emphasize that there are scenarios in which the parameters $\theta_s, m_4$ are unconstrained. As a concrete example, let us consider a Dirac neutrino portal ($m_N=0$ in~(\ref{Lagrangian})). In this case $\nu_s$ can be produced by oscillations with active neutrinos only after $\phi$ has acquired a vacuum expectation value~\cite{Vecchi:2016lty}. Hence, a phase transition occurring below the sterile mass, i.e. $\langle\phi(T)\rangle/m_4(T)\lesssim1$, automatically evades all constraints from BBN and CMB. This particular example requires $y_s\sim1$ and implies $m_A/g_A\sim\langle\phi\rangle\lesssim m_4$, but leaves $\theta_s, m_4, g_A$ as free parameters~\cite{Vecchi:2016lty}. Such regime would extend throughout the ``allowed" green region below the ``Solar Dark MSW" and above the ``IceCube" lines as $m_4$ varies in the range $10^{-2}-10^5$ eV (recall that Fig.~\ref{fig:dMSW} has $m_4=1$ eV).  

The above example does not exhaust all the possibilities, and other relatively unconstrained scenarios exist. The bottom line is that the bounds on the $\theta_s$-link of Fig.~\ref{fig:scheme} arising from Cosmology in general depend critically on additional parameters of the Sterile Sector besides $\theta_s, g_A, m_A$. For this reason we will not consider them any further.

\subsubsection{The $U(1)$ link}

Regarding the Dark Matter-Sterile link described by Eq.~(\ref{Lagrangian}), several signatures can arise. First of all, a coupling between DM and neutrinos ($\propto q_{s}^{2}q_{X}^{2}$) might allow efficient momentum transfer between the relativistic neutrinos and the non-relativistic DM. If this process continues until late times it can prevent the formation of the first gravitationally bound DM structures and come into conflict with the observed small-scale structure of the Universe. Second, there are bounds from the self interaction of DM ($\propto q_{X}^{4}$),  which can modify the halo dynamics and impact small-scale structure. Third, our model introduces an interaction ($\propto \theta_s^2q_{s}^{4}, \theta_s^4q_{s}^{4}$) between active neutrinos and a possible $\nu_s,A'$ component of the cosmic background. In this case we must consider the constraints given by astrophysical neutrinos traveling over large distances (SN 1987A and IceCube neutrinos). Also, loops of active/sterile neutrinos generate effective matter-$A$ interactions ($\propto q_s^2\theta_s^2$). However, these are suppressed by $m_s^2/m_W^2$ and do not lead to significant constraints. Lastly, we can have a modification of neutrino oscillations due to an Asymmetric Dark Matter density ($\propto \theta_s^2q_{s}^{2}q_{X}^{2}$), that is the main subject of this paper.

We now discuss in detail all the cases reported above.

\begin{figure*}[t]
\begin{center}
\includegraphics[width=13cm]{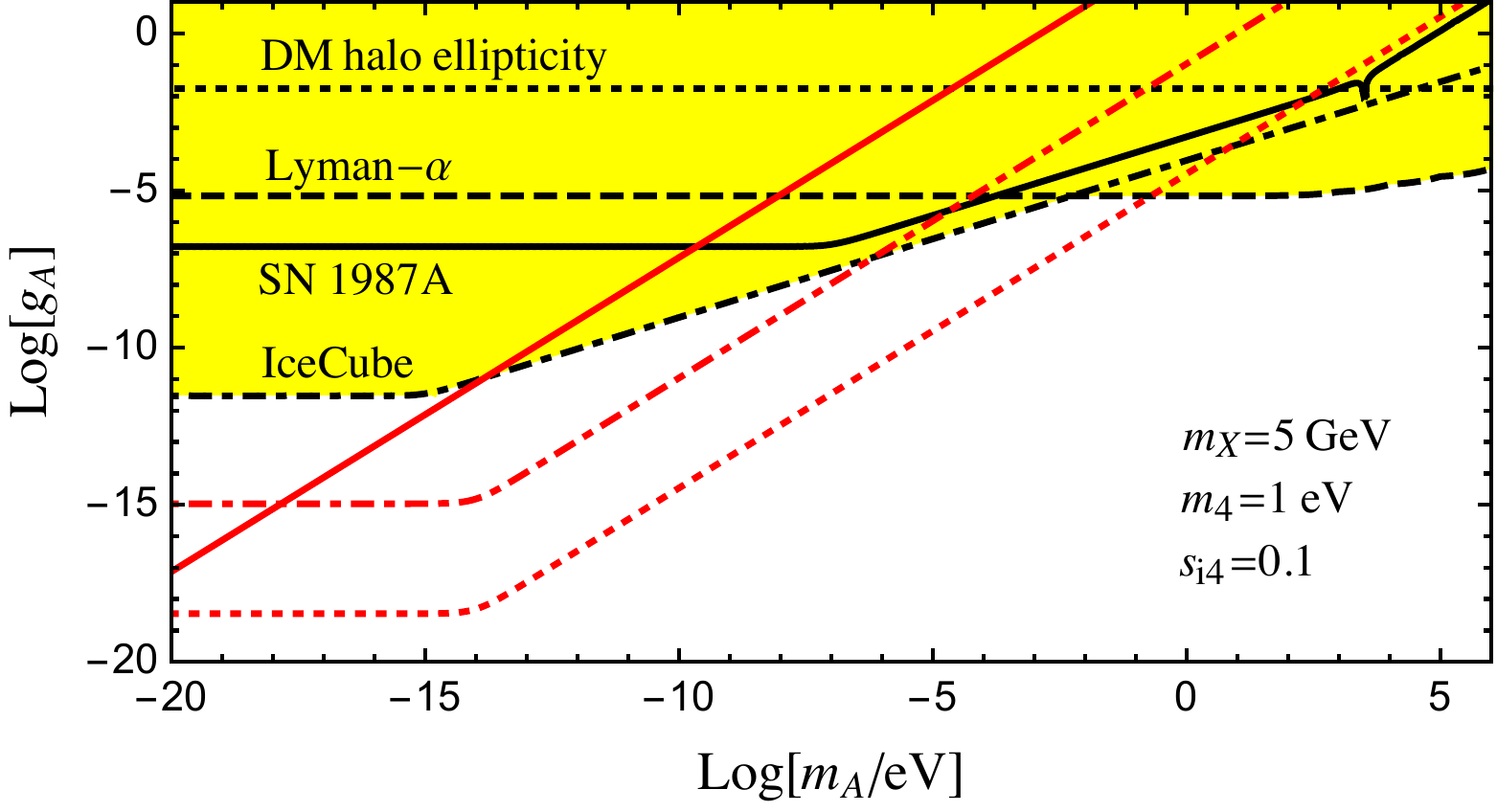}
\caption{The filled yellow region above the black (dotted, dashed, dot dashed, solid) curves is constrained by a variety of considerations. Details can be found in Section~\ref{sec:constr}. For illustrative purposes we have fixed the DM mass to $m_{X} = 5$ GeV, the dark charges to $|q_s|=|q_X|/2=1$, $m_4=1$ eV, and $\sin \theta_s =0.1$. Above the red curves Dark MSW takes place in the Halo (solid) and/or the Sun (dotted and dot-dashed lines) and may thus be probed by neutrino experiments (see the end of Section~\ref{sec:constr} and Section~\ref{sec:DarkMSW}). Note that Solar Dark MSW assumes the existence of interactions between ordinary matter and dark matter beyond those in (\ref{Lagrangian}), that account for a non-vanishing DM population in the sun.}
\label{fig:dMSW}
\end{center}
\end{figure*}

\paragraph{Interactions between light species and Dark Matter} 

The presence of a coupling between light species (active or sterile neutrinos and $A'$) and DM can impact DM physics in interesting ways. In fact, it has been argued that observational evidence may prefer DM-neutrino interactions~\cite{Aarssen:2012fx,Shoemaker:2013tda} and solve the ``missing satellites problem''~\cite{Klypin:1999uc,Moore:1999nt,Kauffmann:1993gv}.

That the DM-neutrino interactions posited here can impact the small-scale structure follows from the fact that these produce a mechanism whereby the relativistic neutrinos can impart a large momentum to DM via elastic scattering, $X \nu \leftrightarrow X \nu$. If this persists until very late times it will prevent the formation of the smallest gravitationally bound DM structures: the so-called proto-halos. Based on the picture of hierarchical structure formation, it is these smallest proto-halos that merged to form larger and larger structures.  A conservative limit from this process is provided by Lyman-$\alpha$ data which indicate that the cutoff in the matter power spectrum induced by this effect cannot be larger than $M_{{\rm cut}} \lesssim 5 \times 10^{10}~M_{\odot}$~\cite{Aarssen:2012fx}.  Attempts to address the missing satellites problem yield cut-offs just below this bound.

Before kinetic decoupling, the DM density perturbations do not grow, but instead undergo ``acoustic oscillations" which efficiently damps the growth of structure~\cite{Boehm:2000gq,Chen:2002yh,Loeb:2005pm,Green:2005fa}. Thus structure cannot grow as long as the momentum-relaxation rate exceeds the Hubble rate. Following \cite{Binder:2016pnr} and earlier references we compute the momentum-relaxation rate via
\begin{equation} 
\gamma(T) = \sum_{i} \frac{g_{i}}{6 m_{X} T} \int_{0}^{\infty} \frac{d^{3}p}{(2\pi)^{3}} f_{i} (1 \pm f_{i}) \frac{p}{\sqrt{p^{2} + m_{i}^{2}}} \int_{-4p^{2}}^{0} dt(-t) \frac{d \sigma_{X + i \rightarrow X + i}}{dt}
\label{eq:gam}
\end{equation} 
where $i$ are the DM's scattering partners. In our case, these scattering partners include both $\nu_s$ and the new gauge boson $A'$, both conservatively assumed to have a thermal distribution. It turns out that the Compton-like scattering process $X + A' \rightarrow X + A'$ sets the strongest limits on the coupling for the range of mediator masses considered here. 

Kinetic decoupling is obtained when the momentum relaxation rate drops below the Hubble rate, so we define the kinetic decoupling temperature as $\gamma (T_{{\rm KD}}) = H(T_{{\rm KD}})$.  The mass of the largest gravitationally bound objects (i.e. the proto-halos) that can form causally is dictated by the mass enclosed in a Hubble volume at $T_{{\rm KD}}$. We then require that this mass not exceed $M_{{\rm cut}}^{{\rm Lyman}-\alpha} = 5 \times 10^{10}~M_{\odot}$, which equivalently can be expressed by requiring that kinetic decoupling occur at temperatures satisfying $T_{{\rm KD}} > 0.15~{\rm keV}$.  This constraint is depicted as the dashed black curve in Fig.~\ref{fig:dMSW}. The line is to be intended as a qualitative indication of the actual bound: a more detailed analysis of structure formation would be necessary to set robust limits. Note that the bound is weaker in realistic scenarios, where the relic density of $A',\nu_s$ is smaller than thermal.

\paragraph{DM self-interactions} 

The $U(1)$ interaction in (\ref{Lagrangian}) also allows for DM-DM scattering with potential impacts on halo ellipticity~\cite{Buote:2002wd,Ackerman:mha,Feng:2009mn}, cluster mergers~\cite{Markevitch:2003at}, as well as the mass profile of dwarf galaxies~\cite{Spergel:1999mh,BoylanKolchin:2011de}. We find that the strongest constraint in our case comes from the observed halo ellipticity, which can be negatively impacted by the presence of overly strong DM self-interactions which drive the halo towards a spherical configuration (see e.g.~\cite{Tulin:2013teo}). This constraint was recently revised to be somewhat less stringent than originally thought~\cite{Agrawal:2016quu}. The latter is shown in the dotted black curve in Fig.~\ref{fig:dMSW}.

\paragraph{Sterile neutrinos in the cosmic background} 

The additional interactions can modify neutrino propagation by the scattering of an incident active neutrino on a relic neutrino (or anti-neutrino), a relic vector boson or the dark matter. By conservatively assuming the relic abundance of steriles is comparable to actives ($100/$cm$^3$) follows that scattering on the sterile neutrino background is the only relevant process. Indeed, under this conservative hypothesis scattering over DM is negligible because of the much smaller density, whereas scattering on vectors is a subleading effect because not enhanced at small momentum transfers.

It has been argued that the apparent isotropy of the IceCube neutrino flux can be used to constrain these processes~\cite{Cherry:2014xra}. Specifically, if the scatterings $\nu_i\overline{\nu_4}\to\nu_4\overline{\nu_4}$ or ${\nu_i}{\nu_4}\to\nu_4{\nu_4}$, with $i=1,2,3$, were sufficiently strong they would erase distant neutrino sources by converting the active neutrinos $\nu_i$ into sterile states, thereby effectively erasing the distant neutrino flux. Then only those neutrino sources at distances much less than the neutrino mean free path (MFP) would be observable. Given that the observable Universe only appears isotropic on scales $\gg 50$ Mpc, we require that the neutrino MFP for $E_\nu\sim$ PeV exceeds 50 Mpc. The resulting constraint is depicted as the dot-dashed black curve in Fig.~\ref{fig:dMSW}. A related though weaker constraint comes from requiring that the MFP for $1-10$ MeV neutrinos be larger than the distance to SN1987A in the Large Magellanic Cloud, as first discussed in~\cite{Kolb:1987qy}. The corresponding bound is shown as a solid black line in Fig.~\ref{fig:dMSW}.

In both lines there is a spike at $m_A^2\sim m_4E_\nu$ due to resonant $\nu_i$ absorption, though for $m_4=1$ eV the resonance in the IceCube curve occurs at $m_A$ higher than shown in our plot. Also, the total cross section $\sigma_{\rm tot}=\sigma(\nu\overline{\nu}\to\nu\overline{\nu})+\sigma(\nu{\nu}\to\nu{\nu})$ for the range of parameters we are interested in scales as $g_A^4/m_A^2$ until the vector mass becomes so small that $A$ appears effectively massless. In the latter regime the force is long range and dominated by the t-channel exchange of the exotic boson. Here we find $\sigma_{\rm tot}\sim g^4/t_{\rm min}\sim g^4s^2/m_4^6$, explaining the plateau in the solid and dot-dashed curves of Fig.~\ref{fig:dMSW}.

We stress that both the IceCube and SN1987A constraints of Fig.~\ref{fig:dMSW} assume equal mixing $\sin \theta_s =0.1$ with all mass eigenstates. If mixing with one of the light mass eigenstates is absent, the latter state would propagate unaffected through the cosmic background and the IceCube and SN1987A constraints depicted in Fig.~\ref{fig:dMSW} would cease to be effective. Similarly to the Lyman-$\alpha$ constraint, these bounds are actually weaker in realistic scenarios with a relic density of $\nu_s$ smaller than thermal.

\paragraph{Neutrino-induced interactions between ordinary matter and $A'_\mu$} 

Loops of active/sterile neutrinos induce an effective $U(1)$ coupling for ordinary matter. For example, at 1-loop one obtains $(g_{\rm eff})_{\alpha\beta}A'_\mu\overline{\ell_\alpha}\gamma^\mu \ell_\beta$, with $\ell_\alpha$ a charged lepton. The relevant diagrams are finite and involve $W^\pm$ exchange as well as (at least) two insertions of the mixing. The typical virtual momenta in the loop is of order $m_W$ and the effective coupling parametrically scales as $g_{\rm eff}\sim g_A\sin^2\theta_s({g^2}/{16\pi^2})({m_4^2}/{m_W^2})$. For $m_4=1$ eV this is so small that even the stringent constraints from fifth force experiments are satisfied with $g_A$ of order unity. For all practical purposes we can safely ignore such effective coupling.

\paragraph{New matter effects in neutrino propagation} 

A vast region of parameter space in Fig.~\ref{fig:dMSW} is left unconstrained by the bounds discussed above (yellow region). In this paper we point out that at small couplings and mediator masses the main impact of the new DM coupling may show up as a modification of the propagation of active neutrinos in a DM medium. We refer to this as the Dark MSW effect.

The coherent scattering of active neutrinos, via oscillations into $\nu_s$, is affected by DM already at tree level. In the limit of zero average velocity of the DM, $\langle\overline{X}\gamma^\mu X\rangle=n_X\delta^{\mu0}$, the effect is described by a DM potential for the sterile component:
\ba\label{potentialGen}
V_{\rm eff}=q_sq_X\frac{g_A^2}{\partial^2+m_A^2}n_X.
\ea
$n_X$ is the DM charge density and is nonvanishing only if the density of DM particles is different from that of anti-particles, a condition that is naturally met in scenarios with a primordial DM asymmetry. (Without loss of generality we conventionally take $n_X>0$.)

In Eq.~(\ref{potentialGen}) we see that the mass of the mediator, $m_A$, sets a length scale to be compared to the gradient of $n_X$. ${V}_{\rm eff}$ describes a Coulomb potential $\sim1/r$ in the limit $m^2_A\ll(\partial^2n_X)/n_X$, whereas for larger $m_A$, Eq.~(\ref{potentialGen}) reduces to a familiar contact interaction $V_{\rm eff}=G_Xn_X$ with
\ba\label{Gx}
G_X=q_sq_X\frac{g_A^2}{m_A^2}.
\ea
Because for the systems we are interested in $(\partial^2n_X)/n_X$ vary on very large length scales -- say of order $>10^4$ km -- we can safely assume the contact limit.

A key aspect of Dark MSW is that it depends on two unknown quantities: the DM charge density $n_X$ and the mediator's coupling $G_X$. This has important consequences. For example, no large DM densities are required to generate an appreciable effect because the unknown parameter $G_X$ can be many orders of magnitude larger than the Fermi coupling whenever $m_A\ll100$ GeV. As a prototypical example, $|q_s|\sim|q_X|\sim1$ and $m_A/g_A=1$ eV gives $G_X/G_F\sim10^{23}$. Such impressive numbers are not excluded by the bounds of the previous subsections (in fact they are motivated by the discussion in Section~\ref{motivated}), and can easily compensate an otherwise small $n_X$.

The most familiar environments with DM are our galaxy halo, and possibly the sun and the earth. The effect of the galactic DM halo on neutrino propagation has been previously considered in~\cite{Horvat:1998ym,Lunardini:2000swa,Miranda:2013wla,deSalas:2016svi}. Such phenomenon may occur in virtually all oscillation experiments where $n_XG_XE\gtrsim\Delta m^2$. Taking $n_X{m_X}\sim0.4~{\rm GeV}/{\rm cm}^3$ and $\Delta m^2\sim\Delta m_{31}^2$ as typical values, such condition becomes $G_XE/m_X\gtrsim10^{20}/$GeV$^2$. For Asymmetric DM $m_X$ naturally lies in the GeV range~\cite{Kaplan:2009ag}, so the above requirement becomes roughly $G_X\gtrsim10^{23}/$GeV$^2({\rm MeV}/E)$. Note that for $E\sim1$ MeV, this may not be possible in the framework of Eq.~(\ref{Lagrangian}) because it would require $\langle\phi\rangle\lesssim10^{-2}$ eV, which is not consistent with a $m_4\sim y_s\langle\phi\rangle$ in the eV range. This argument illustrates that effects due to the DM halo are typically visible in the oscillation of high energy neutrinos, that are unfortunately poorly understood.

In this paper we note that if a fraction of DM has accumulated in the sun due to a non-vanishing $\sigma_{nX}$, a much smaller $G_X$ is sufficient to have a non-trivial impact of the DM potential Eq.~(\ref{potentialGen}) on solar neutrinos. This possibility is particularly exciting because solar data are quite well understood and can provide interesting probes of our scenario. We turn to a study of Solar Dark MSW in Section~\ref{sec:DarkMSW}.

\paragraph{Supernovae explosion} 

Before analyzing Solar Dark MSW in detail it is worth to emphasize that also Supernovae explosions can be affected by the presence of exotic neutrinos as well as the new gauge interaction. However, as opposed to analogous effects in the sun, it is extremely difficult to precisely quantify the impact on Supernovae neutrinos because our current understanding of their physics is not fully satisfactory even in the standard 3-neutrino framework. For this reason no ``Supernovae constraint" is shown in Fig.~\ref{fig:dMSW}. Yet, for completeness here we present a qualitative discussion of the relevant physics and refer the interested reader to the appropriate literature.

Following the same logic of Section~\ref{intro}, one estimates that DM may be captured by a Supernovae and confine in a region of radius $\sim3$ km, well within the neutrino sphere. Now, for small DM potentials the steriles are expected to behave as truly-sterile neutrinos, so we can use the results of~\cite{Mikheev:1987hv,Nunokawa:1997ct,Abazajian:2001nj,Cirelli:2004cz,Tamborra:2011is} and references therein. There it was argued that due to the sizable ordinary-matter interactions felt by the active neutrinos, the effective active/sterile mixing inside the star is very small and only a tiny fraction of steriles can be produced. Conversions may however take place far outside, where various resonances occur. The latter can trigger neutrino and anti-neutrino transitions into steriles as well as convert steriles back into actives. The resulting survival probability for electron neutrinos and anti-neutrinos is sufficiently large that the few events from the SN1987A appear to imply no constraint on our benchmark scenario with $\Delta m_{41}^2=1$ eV$^2$, $\sin\theta_s=0.1$~\cite{Nunokawa:1997ct,Abazajian:2001nj,Cirelli:2004cz,Tamborra:2011is}. 

The physics is a bit different for large DM potentials. In this case it may be possible to have resonant conversion of actives into steriles already inside the DM core, though only for either neutrinos or anti-neutrinos. Given that the DM density extends up to radii of the order of a third of the neutrino sphere, in the worst case scenario we expect that at most a fraction of order unity of the active neutrinos produced within the core are converted into steriles. Hence, even accepting that the latter are completely lost, SN1987A cannot be used to constrain our scenario in this case either. In reality, as described in the previous paragraph, resonances taking place away from the core alter significantly the flavor content of the neutrinos emitted by the star and potentially further re-populate the active species. 

Our conclusion is that current and future observations of Supernovae explosions may be used to constrain our framework only once a robust understanding of its physics is developed.

\section{Solar Dark MSW}
\label{sec:DarkMSW}

In this section we study the impact that a non-vanishing asymmetric DM density in the sun could have on solar neutrino, a phenomenon that we call Solar Dark MSW. 

In Sec.~\ref{intro} we estimate the DM density profile $n_X$ in an asymmetric DM model and briefly mention the qualitative features that distinguish the Solar Dark MSW phenomenon from other exotic modifications of solar neutrino physics. The relevant effective Hamiltonian is introduced in Section~\ref{surv}, where we also provide a useful analytic expression of the neutrino survival probability. A more quantitative assessment of Solar Dark MSW is provided by our numerical analysis of Section~\ref{sec:constraints}.

\subsection{Dark Matter in the Sun}
\label{intro}

As long as DM has nonvanishing interactions with ordinary matter ($\sigma_{nX}$ is non-zero in asymmetric dark matter models because visible-dark interactions are necessary to transfer the asymmetry from one sector to the other), $X$ may collide with the stellar medium, lose energy and get trapped by the star's gravitational pull~\cite{Gould:1987ir}. Once captured, DM thermalizes and distributes within a thermal region of density
\ba\label{DMdensity}
n_X(r)= \frac{N_X}{r_X^3\pi^{3/2}}e^{-r^2/r_X^2},~~~~~~~~~r_X=\sqrt{\frac{3T_{\odot}}{2\pi G_N\rho_{\odot} m_X}}\sim0.05~\sqrt{\frac{5~{\rm GeV}}{m_X}}~R_\odot,
\ea
inside the host star. Here we normalized the density as $\int d^3r~n_X=N_X$, with $N_X$ the total number of trapped DM particles (in our scenario the population of DM anti-particles is negligible today). In deriving Eq.~(\ref{DMdensity}) we took a constant matter density $\rho_{\odot}\sim150$ g/cm$^3$ and $T_\odot\sim10^7$ Kelvin, which turns out to be a good approximation since for the cases we are interested in $r_X\ll R_\odot$. Note that in the sun $1/r_X\sim10^{-14}~{\rm eV}\sqrt{m_{X}/5~{\rm GeV}}$, so as anticipated around Eq.~(\ref{potentialGen}), as long as $m_{A}$ is larger than this value the potential is well approximated by $V_{\rm eff}  \simeq G_{X} n_{X}$. We will focus on this case throughout the paper.

In order to estimate $N_X$ we assume that DM is sufficiently heavy that evaporation can be ignored, $m_X\gtrsim4$ GeV (see e.g.~\cite{Busoni:2013kaa}), and that DM self-interactions are not important. In this case the evolution of DM in the Sun is straightforward, $N_{X}(t) = C t$, and we find $N_X($now$)\sim10^{34}\left({\sigma_{nX}}/{10^{-45}~{\rm cm}^2}\right)$. From Eq.~(\ref{DMdensity}) it follows that this corresponds to a density of order ($m_X=5$ GeV)
\ba
\label{eq:est}
\frac{n_X}{n_e}\sim10^{-21}\left(\frac{\sigma_{nX}}{10^{-45}~{\rm cm}^2}\right)~~~~~~~~({\rm Sun}).
\ea
Compared to the typical electron density $n_e\sim100~N_A/$cm$^3$, this is a very tiny number.~\footnote{For such values solar physics, and in particular neutrino production, is not appreciably modified by the presence of DM and the standard solar models can be employed.} However, it can easily be compensated by a large $G_X$. Specifically, the region where the ``Solar Dark MSW" effect takes place, identified as $n_XG_X\sim n_eG_F$ (see section~\ref{sec:constraints}), lies above the red curves in Fig.~\ref{fig:dMSW}. 
In the figure we used Eq.~(\ref{eq:est}) with $\sigma_{nX}=10^{-45}$ cm$^2$ (dot-dashed) and $\sigma_{nX}=10^{-38}$ cm$^2$ (dotted) and included the transition to the regime where the potential is Coulomb-like. We conclude that solar neutrinos can probe a considerable region of the parameter space. Moreover, as anticipated in the previous section, for DM masses in the GeV range the condition $G_Xn_XE\sim\Delta m^2$ necessary to have a halo effect at $E\sim10$ MeV requires much larger couplings -- see the region above the solid red curve of Fig.~\ref{fig:dMSW}.

A calculation similar to the one leading to Eq.~(\ref{eq:est}) suggests that the earth itself may contain a DM density $n_X/n_e\sim10^{-31}\left({\sigma_{nX}}/{10^{-45}~{\rm cm}^2}\right)$, many orders of magnitude smaller than in the sun. For $G_X$ large enough Dark MSW would thus be visible in the passage through the earth's core, as well. However, for such enormous values of $G_X$ solar data would already provide considerable constraints (as we will see below). We can therefore safely ignore Earth's (and Halo's) effects in what follows.

\begin{figure}[t]
\begin{center}
\includegraphics[width=9.5cm]{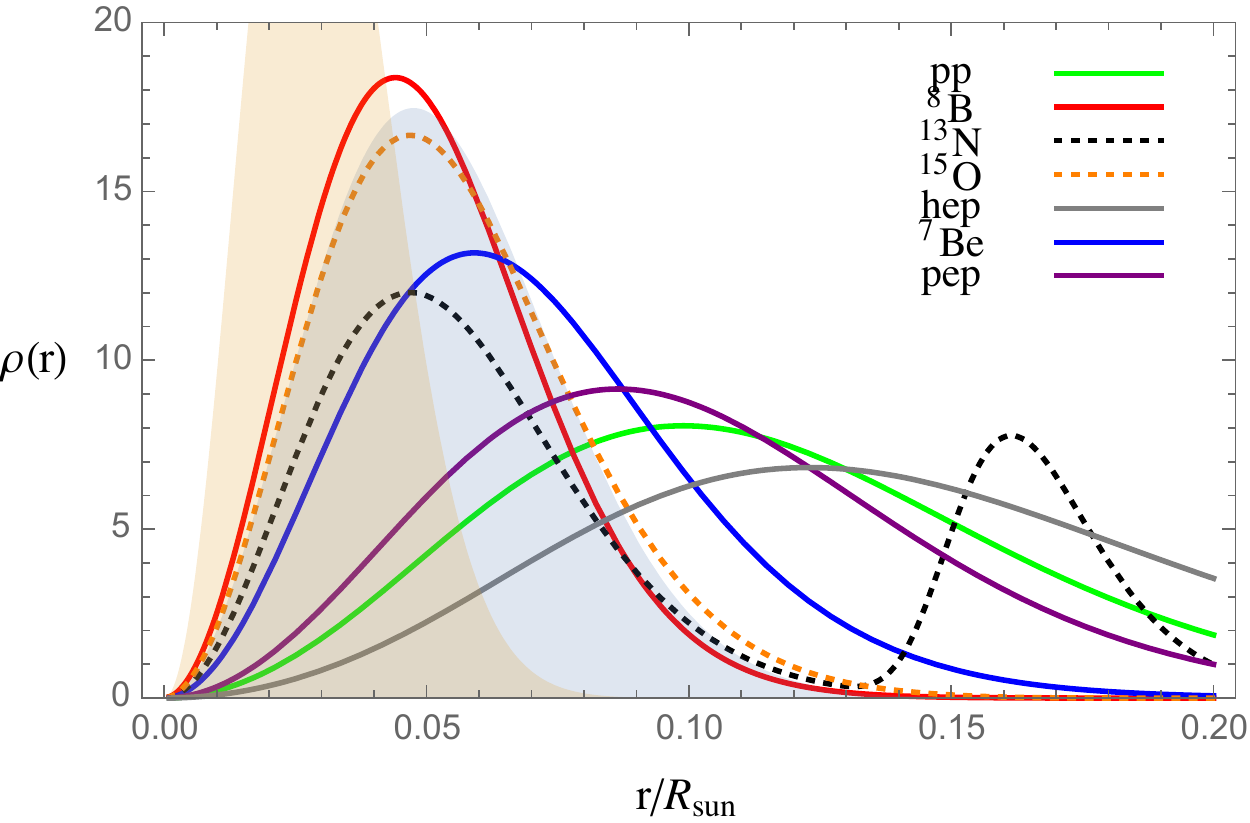}
\caption{Production point (in units of the sun radius and normalized such that $\int dr\rho_i=1$) for various neutrino species. We also show a typical thermal DM density distribution $\rho_X(r)=4\pi(r^2/r_X^3)e^{-r^2/r_X^2}$ for $m_X=5$ GeV (filled yellow region) and $m_X=15$ GeV (filled orange region), see (\ref{DMdensity}).
}\label{densities}
\end{center}
\end{figure}

Now, Eq.~(\ref{DMdensity}) illustrates a key aspect of solar Dark MSW: neutrino species are affected differently by DM. Indeed, asymmetric DM models usually require masses in the range $m_X\sim 5-15$ GeV~\cite{Kaplan:2009ag}. From Eq.~(\ref{DMdensity}) we see that with these values the DM density peaks at around $3-5\%$ of the solar radius, precisely where the most energetic neutrinos are produced, see Fig.~\ref{densities}. This implies that $^8$B and CNO neutrinos (mostly $^{13}$N, $^{15}$O) are the species more affected by the DM potential in Eq.~(\ref{potentialGen}). The numerical study of Sec.~\ref{sec:constraints} will explicitly confirm this expectation. Note that for smaller $m_X$ pp-neutrinos would start to feel the DM potential more significantly; however, in this regime evaporation effects would significantly impact the capture rate and basically prevent the formation of a DM core in the sun. In other words, the statement that solar Dark MSW impacts mostly $^8$B and CNO neutrinos is robust.

\subsection{The Survival Probability}
\label{surv}

We now discuss how the new dark matter potential Eq.~(\ref{potentialGen}) with $n_X$ given in Eq.~(\ref{DMdensity}) impacts solar neutrinos. We consider for simplicity the case in which a single sterile neutrino $\nu_s$ mixes with the active ones. The analysis can be straightforwardly generalized.

The propagation of neutrinos in medium is described by the following Hamiltonian:
\ba\label{hamilton}
H=\frac{1}{2E}U\left(
\begin{array}{cccc}
0 &  & & \\
 & \Delta m_{21}^2 & & \\
 & & \Delta m_{31}^2 &\\
 & & & \Delta m_{41}^2
\end{array}
\right)U^\dagger
+
\left(
\begin{array}{cccc}
V_{\rm CC} &  & & \\
 & 0 & &  \\
 & & 0 &\\
 & &  & V_s
\end{array}
\right),~~~~V_s=V_{\rm eff}-V_{\rm NC}
\ea
where $\Delta m^2_{i1}=m_i^2-m_1^2$, $V_{\rm CC}=\sqrt{2}G_Fn_e(r)$, $V_{\rm NC}=\sqrt{2}G_F\frac{1}{2}n_n(r)$, and for later convenience we write $V_{\rm eff}=\sqrt{2}\xi G_F n_e(0)e^{-r^2/r_X^2}$, where we introduced the parameter
\ba\label{Xi}
\xi\equiv\frac{G_Xn_X(0)}{\sqrt{2}G_Fn_e(0)}.
\ea
The vacuum mixing is controlled by $U=U_{34}R_{24}U_{14}R_{23}U_{13}R_{12}$,~\footnote{We describe an alternative parameterization (used to check our numerical results) in Appendix \ref{sec:alt}.} where all angles and phases vary within $0\leq\theta_{ij}\leq\pi/2$ and $0\leq\delta_{ij}<2\pi$. As usual, for anti-neutrinos the potentials are reversed.

For the range of parameters we are interested in, the propagation in the sun is completely adiabatic. The reason is that the DM potential varies on scales $r_X$ comparable to the sun radius, see Eq.~(\ref{DMdensity}), so the non-adiabaticity parameter on resonance is similar in size to the standard case. Furthermore, given the huge distances involved, the neutrino flux reaching the earth is highly incoherent (provided $\Delta m^2_{i1}\gg10^{-8}$ eV$^2$, which we assume). Under these assumptions the probability that an electron neutrino produced at the radial distance $r$ from the center of the Sun be detected on earth as an electron neutrino (aka survival probability) becomes 
\ba\label{PeeDef}
P_{ee, {\rm day}}(r,E)=\sum_i |U'_{ei}(r,E)|^2|U_{ei}|^2=\sum_iP_{ei}^\odot P_{ie}^{(0)},
\ea
where $U'$ is the mixing matrix in the sun. Our numerical code directly implements this simplified expression.

During night, neutrinos traverse the Earth and the vacuum oscillation probability $P_{ie}^{(0)}$ should be replaced with the corresponding probability on Earth, $P_{ee,{\rm night}}=\sum_iP_{ei}^\odot P_{ie}^\oplus$. This latter effect cannot ignore oscillations and must be treated carefully. The calculation of $P_{ie}^\oplus$ follows the approach proposed in~\cite{Lisi:1997yc}: we determined the evolution operator for neutrinos crossing the Earth through a second order Magnus expansion \cite{Magnus:1954zz}; we approximated the Earth electron density, as reported by the PREM model \cite{Dziewonski:1981xy}, with a bi-quadratic polynomial, which allows to perform analytically the integrals required by the expansion; finally, we performed the average of $P_{ie}^\oplus$ over the nadir angles using the weighting function for SNO reported in Fig. 3 and Table II of \cite{Lisi:1997yc}.

\subsubsection{Analytic approximation for $P_{ee,{\rm day}}$}
\label{sec:analyt}

There are several resonances that can potentially affect $P_{ee}$. Besides the standard one, there are two new ones at $V_sE\sim\Delta m_{31}^2, \Delta m_{41}^2$. Note that $V_sE\sim\Delta m_{21}^2$ represents just a modification of the standard MSW resonance. Also, because we focus on scenarios with $\Delta m_{41}^2\sim1$ eV$^2$, it turns out that the resonance with the heavier state is never relevant. In practice, therefore, the only new resonance we will encounter is controlled by $V_sE\sim\Delta m_{31}^2$.

Away from such resonance it is possible to approximate Eq.~(\ref{hamilton}) with a simple 2 by 2 problem~\cite{Kuo:1986sk}. There are two reasons to carry out such an approximation. First, it allows us to analytically appreciate the main features of our system. Secondarily, it provides a non-trivial crosscheck of the numerical analysis of section~\ref{sec:constraints}.

It is convenient to approach this problem in a new basis defined by the rotation $V=UR^t_{12}$. In the new basis the Hamiltonian reads $H'=V^\dagger HV$. Assuming $s_{i4}^2V_s<\Delta m^2_{31}$, the mixing between the heavy states of ``energy" $\sim\Delta m_{31,41}^2$ and the light states is suppressed. Up to these corrections we can retain the upper 2 by 2 terms in $H'$, that are:
\ba\label{eff}
H'_{2\times2}&=&\left(
\begin{array}{cc}
-\Delta \cos2\theta_{12}+V_x & \Delta \sin2\theta_{12}+V_y \\
\Delta \sin{2\theta_{12}}+V_y^* & \Delta \cos2\theta_{12} - V_x
\end{array}
\right)
\ea
where $\Delta=\Delta m_{21}^2/4E$, $V_x=\frac{1}{2}\left[V_{\rm CC}c_{13}^2c_{14}^2+V_s\left(|A|^2-|B|^2\right)\right]$, $V_y=V_sAB$, and
\ba
A&=&e^{-i\delta_{14}}c_{13}c_{24}c_{34}s_{14}-e^{-i\delta_{13}}s_{13}\left(c_{34}s_{23}s_{24}+e^{-i\delta_{34}}c_{23}s_{34}\right),\\\no
B&=&c_{23}c_{34}s_{24}-e^{i\delta_{34}}s_{23}s_{34}.
\ea
(Here we employ the common abbreviations $s_{ij}^2\equiv\sin^2\theta_{ij}$ and $c_{ij}^2\equiv\cos^2\theta_{ij}$.) Importantly, sizable off-diagonal effects $|V_y/V_x|\gtrsim1$ are possible only when at least two exotic angles are nonzero. 

Up to corrections of order $\mathcal{O}(s_{i4}^2V_s/\Delta m^2_{31})$, the 4 by 4 matter Hamiltonian $H'$ is diagonalized by a complex rotation along the 1-2 directions, $U_{12m}$. Its phase $\phi$ is determined by ${\rm Arg}(\Delta \sin{2\theta_{12}}+V_y)=\phi$ while the angle is given by, 
\ba\label{analtheta}
\cos2\theta_m&=&\frac{\Delta\cos2\theta_{12}-V_x}{\sqrt{|\Delta\sin2\theta_{12}+V_y|^2+(\Delta\cos2\theta_{12}-V_x)^2}}.
\ea
Finally, including all angles and phases, and recalling that the matrix that diagonalizes the original Hamiltonian is approximately given by $U'\approx VU_{12m}$, Eq.~(\ref{PeeDef}) reduces to
\ba\label{PeeApp}
P_{ee,{\rm day}}&=&c_{13}^4c_{14}^4\frac{1}{2}\left(1+\cos2\theta_{12}\cos2\theta_{m}\right)+s_{13}^4c_{14}^4+s_{14}^4\\\no
&+&{\cal O}(s_{i4}^2V_sE/\Delta m_{31}^2).
\ea
Note the strong degeneracy of $s_{24,34},\delta_{13,14,34}$, that are entirely described by $|A|,|B|,{\rm Arg}(AB)$. On the other hand, $s_{14}$ explicitly appears in Eq.~(\ref{PeeApp}).

\subsection{Constraints from solar neutrino data}
\label{sec:constraints}

In this section we present a numerical investigation of Solar Dark MSW. Our primary aim is to quantitatively discuss its distinctive features, as well as assess its compatibility with current solar data. Rather than performing a scan of the full parameter space, as a first look at Solar Dark MSW we find it more instructive to focus around the region $s_{14}^2=0.027$, $s_{24}^2=0.014$, $\Delta m_{41}^2=1.6~{\rm eV}^2$, motivated by the SBL anomalies~\cite{Gariazzo:2015rra}. A more thorough investigation is left for future work.

We focus on the constraints arising from the $^8$B, pep and $^7$Be solar neutrino data. Neutrinos from the pp-chain have a modest effect on our results --- in particular, they do not qualitatively change our contours compared to a SNO-only analysis --- and can therefore be neglected. This is a consequence of the fact that Dark MSW does not impact neutrinos produced away from the core, as anticipated when discussing Fig.~\ref{densities}. $^8$B neutrinos, by far the most crucially affected in our framework, are constrained by SNO~\cite{Aharmim:2011vm} and SK data~\cite{Abe:2016nxk}. The SNO collaboration provides a fit of their data assuming a polynomial form for $P_{ee}$ as well as the day night asymmetry. This approach is extremely useful when testing new physics models like ours, and spares us from a full re-analysis of their data. SK presents a similar analysis for $P_{ee}$ but not for the day-night asymmetry. We therefore opted to use SNO; the inclusion of SK data is not expected to change our results qualitatively.

We define the $\chi^2$ as the sum of three terms:
\ba\label{chi}
\chi^2({\bf{p}})=\chi^2_{\rm SNO}({\bf{p}})+\chi^2_{\rm Be7}({\bf{p}})+\chi^2_{\rm pep}({\bf{p}})\ ,
\label{chi2}
\ea
where $\bf{p}$ represents a generic point of the parameter space, which includes both standard and exotic parameters. In Appendix~\ref{sec:chi} we describe in detail how the $\chi^{2}$ in Eq.~(\ref{chi}) is determined. The exotic parameters in $\bf{p}$ include $\Delta m_{41}^2$, $\xi$, $r_X$, $s_{14,24,34}^2$, $\delta_{13,14,34}$, where we recall that $s_{i4}^2\equiv\sin^2\theta_{i4}$. For illustrative purposes we take $r_X=0.05~R_\odot$ throughout our analysis --- corresponding roughly to $m_X=5$ GeV, see Eq.~(\ref{DMdensity}), a typical expectation in asymmetric DM scenarios. Our results do not change qualitatively if we assume smaller values of $r_X$, whereas larger values are not relevant, as discussed at the end of Section~\ref{intro}. Furthermore, it turns out that $\Delta m_{41}^2$ has little or no impact as long as it is $\gg|\Delta m_{31}^2|$ (we always assumed normal hierarchy for simplicity). For this reason we decided to fix $\Delta m_{41}^2=1.6$ eV$^2$ in all plots, a value favored by short baseline data \cite{Gariazzo:2015rra}, though the reader should be aware that our numerical analysis applies also to heavier steriles. Similarly, the phases $\delta_{13,14,34}$ do not modify our conclusions qualitatively and have been set to zero. Within this simplified picture the relevant exotic parameters reduce to 
\ba
\xi,~~~~s^2_{14,24,34}. 
\ea
The result of our numerical scan is summarized in Figures~\ref{solarFit}, \ref{idea}, \ref{34GX}. The 1, 2, 3$\sigma$ regions correspond respectively to $|\chi^2-\chi^2_{\rm best~fit}|<2.30, 6.3 ,11.8$ (assuming 2 degrees of freedom), and the best fit is identified by a dot.

\subsubsection{Discussion}

\begin{figure}[t]
\begin{center}
\includegraphics[width=8cm]{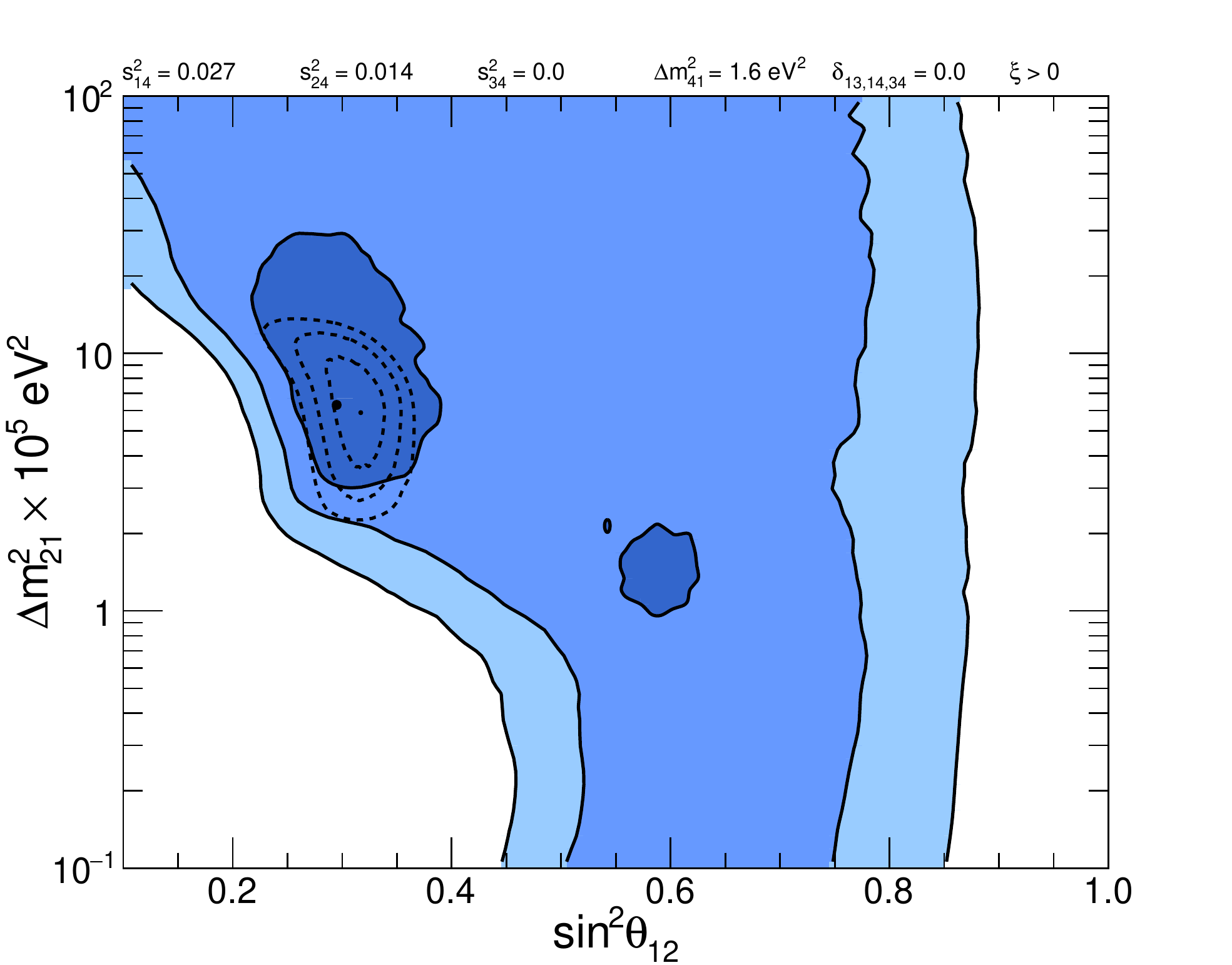}~~\includegraphics[width=8cm]{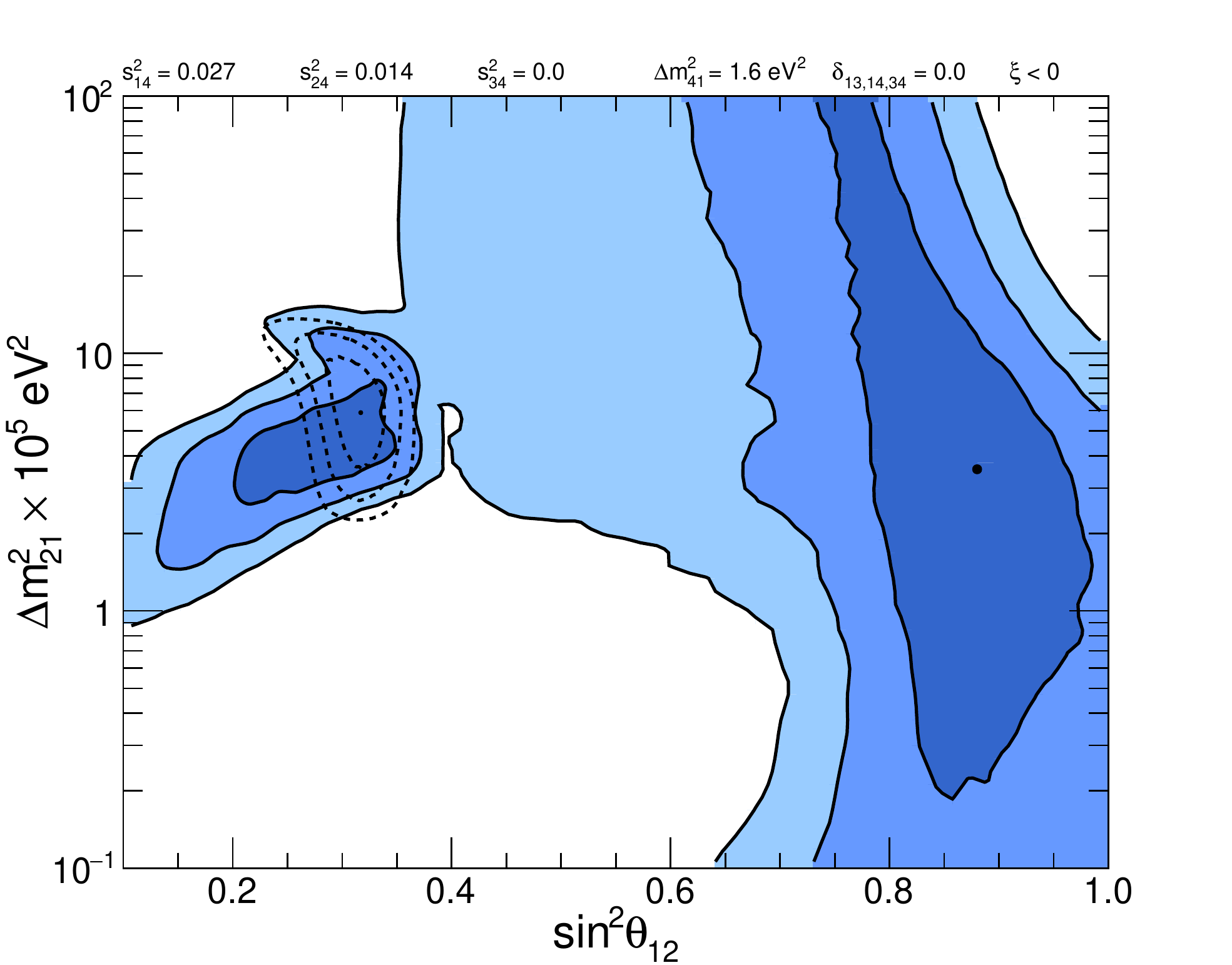}
\caption{LMA and Dark-LMA solutions. Here $s_{14}=0.027$, $s_{24}=0.014$, $s_{34}=0$, $\delta_{13,14,34}=0$, $\Delta 
m_{41}^2=1.6$ eV$^2$, and we scanned over $\Delta m_{21}^2,s_{12}, \xi>0$ (left) 
or $\xi<0$ (right). The colors represent allowed regions for $\left|\chi^2-\chi^2_{\text{best fit}}\right|$ = 2.3, 6.3, 11.8 (corresponding to $1, 2, 3 \sigma$ and 2 degrees of freedom) from darker to lighter. Overlaid in dashed we show the corresponding regions in the Standard 
Model. 
}\label{solarFit}
\end{center}
\end{figure}

Figures~\ref{solarFit} show a scan in $\Delta m^2_{21},s^2_{12},\xi$ with the other parameters held fixed at $s^2_{14}=0.027$, $s^2_{24}=0.014$, $s^2_{34}=0$, $\delta_{13,14,34}=0$, after having marginalized over $\xi$ (the values of $s^2_{14,24}$ are taken from \cite{Gariazzo:2015rra}). These figures reveal two important features. First, when $\xi>0$ the only-solar best fit is naturally pushed towards the KamLAND region \cite{Decowski:2016axc}. Second, in general there exists a Dark-LMA solution, similar to the one found in the context of NSI~\cite{Miranda:2004nb}. Once KamLAND is taken into account, and for the parameters we considered in our simulation, the only Dark-LMA solution that survives is at $\xi<0$ (and 1$\sigma$), however.

\begin{figure}[h]
\begin{center}
\includegraphics[width=8cm]{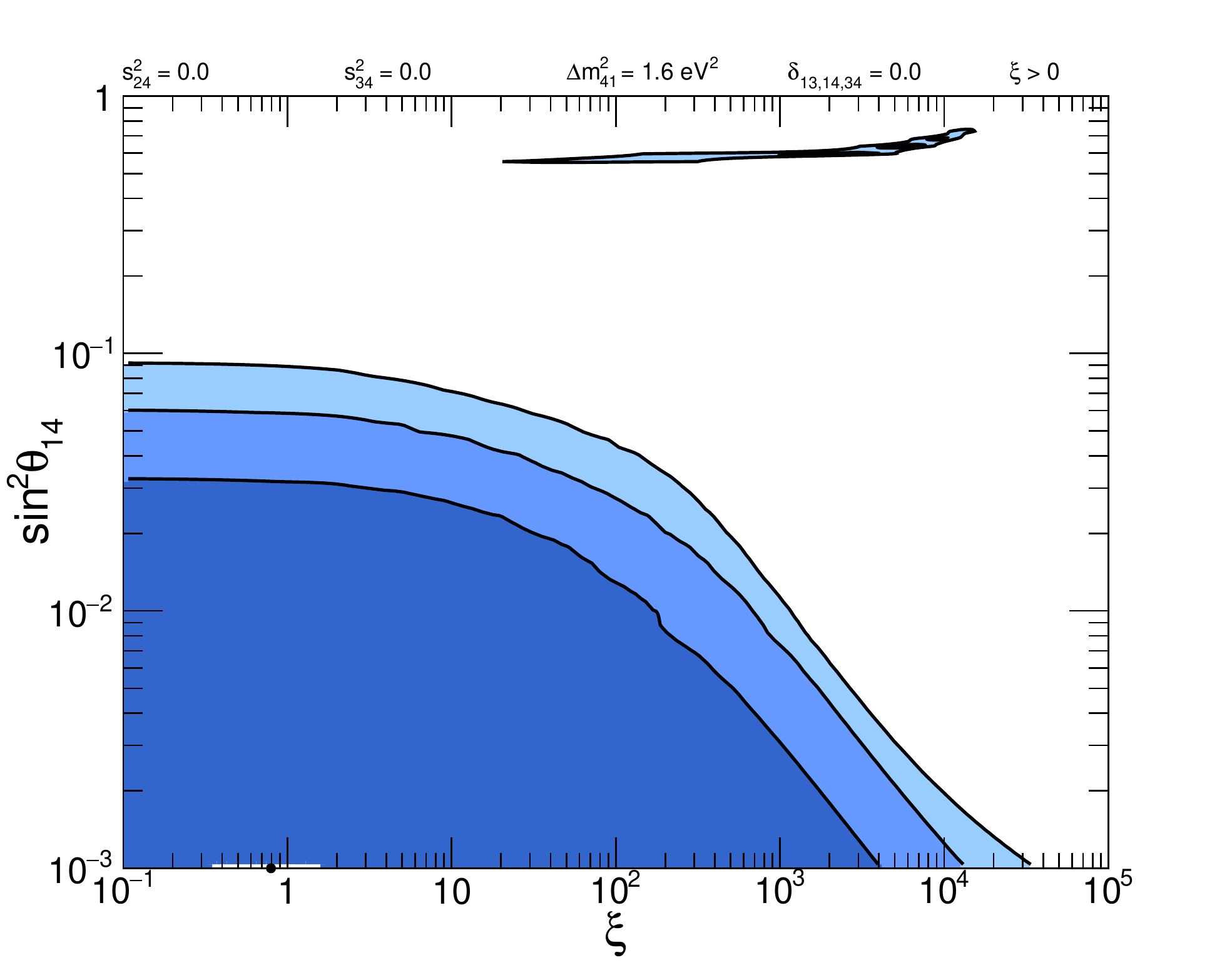}~~\includegraphics[width=8cm]{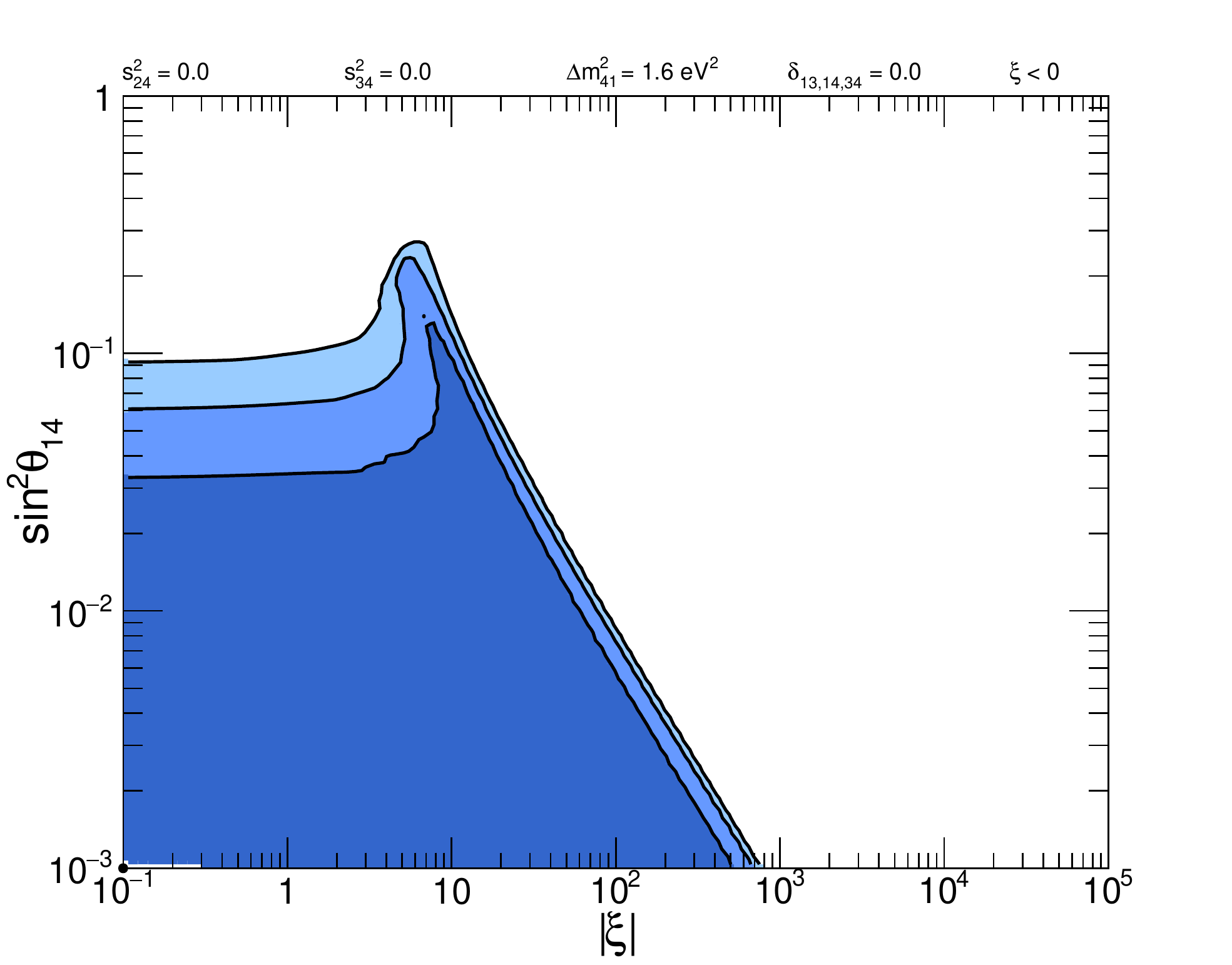}
\includegraphics[width=8cm]{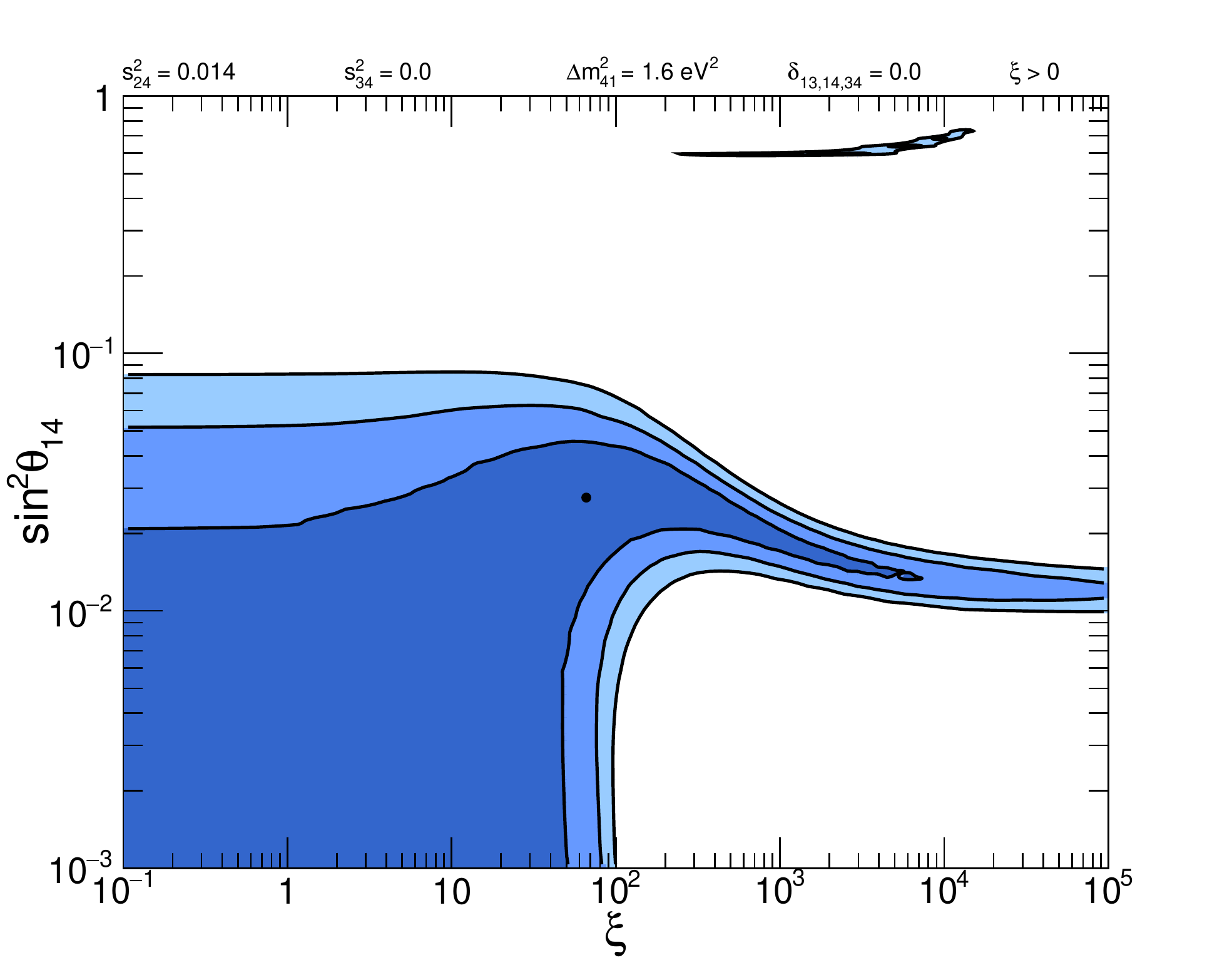}~~\includegraphics[width=8cm]{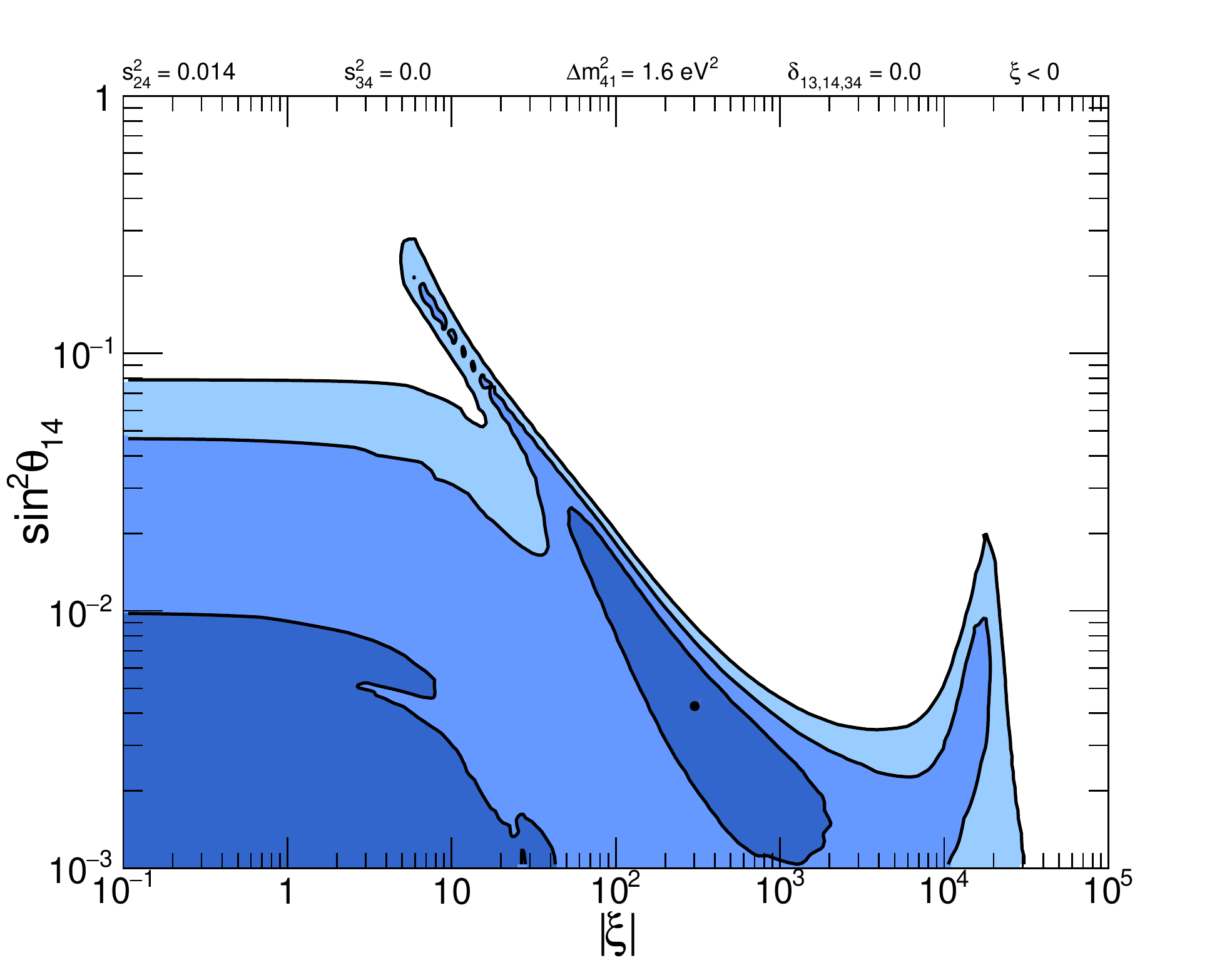}
\caption{Allowed $s_{14}, \xi$ contour regions with $\xi>0$ (left) and $\xi<0$ (right). Here $s_{34}=0=\delta_{13,14,34}$ and $s_{24}=0$ (upper plots), $s_{24}=0.014$ (lower plots). The same confidence
levels reported in Fig. \ref{solarFit} are used.
}\label{idea}
\end{center}
\end{figure}

In the upper part of Fig.~\ref{idea} we present a scan over $s^2_{14},\xi$ with $s^2_{24,34}=0$ and for $\xi>0,<0$. First, note an upper bound of order $s_{14}^2<0.1$ on the exotic angle. This can be understood directly from Eq.~(\ref{PeeApp}), where we see that an otherwise larger $s^2_{14}$ would suppress $P_{ee}$ too much in the vacuum-dominated region. Such suppression cannot be compensated by the other parameters. The other boundary of the allowed region is roughly identified by $|\xi| s_{14}^2\lesssim{\cal O}(10)$. In other words, the new potential cannot be too large compared to the standard one. This restriction is relaxed when more angles are switched on, as we discuss next. The lower part of Fig.~\ref{idea} --- the ``elephant'' ($\xi>0$) and ``rhino'' ($\xi<0$) plots --- shows a scan of $s^2_{14}, \xi$ with $s^2_{24}=0.014, s^2_{34}=0$. We again see the cut in the region $s_{14}^2<0.1$, as in the upper plots. However, the other boundary has changed qualitatively. In particular, part of the domain at large coupling, where according to our discussion in Section~\ref{sec:analyt} resonances involving $\nu_3$ can take place, is now allowed. This region is characterized by abrupt changes in $P_{ee}$ and correspond to the elephant's trunk and the frontal horn of the rhino, where $s_{14}^2|\xi|\sim\Delta m_{31}^2/\Delta m_{21}^2\sim30$ or larger. We conclude that there are interesting points in the parameter space where the problem is not well-approximated by a simple two flavor description.

As a curiosity, we observe that the absolute best fit for $s_{14}^2$ at $\xi>0$ is remarkably close to the one reported in \cite{Gariazzo:2015rra}, although the width of the $1\sigma$ region covers more than one order of magnitude in $s_{14}^2$.

\begin{figure}[h]
\begin{center}
\includegraphics[width=8cm]{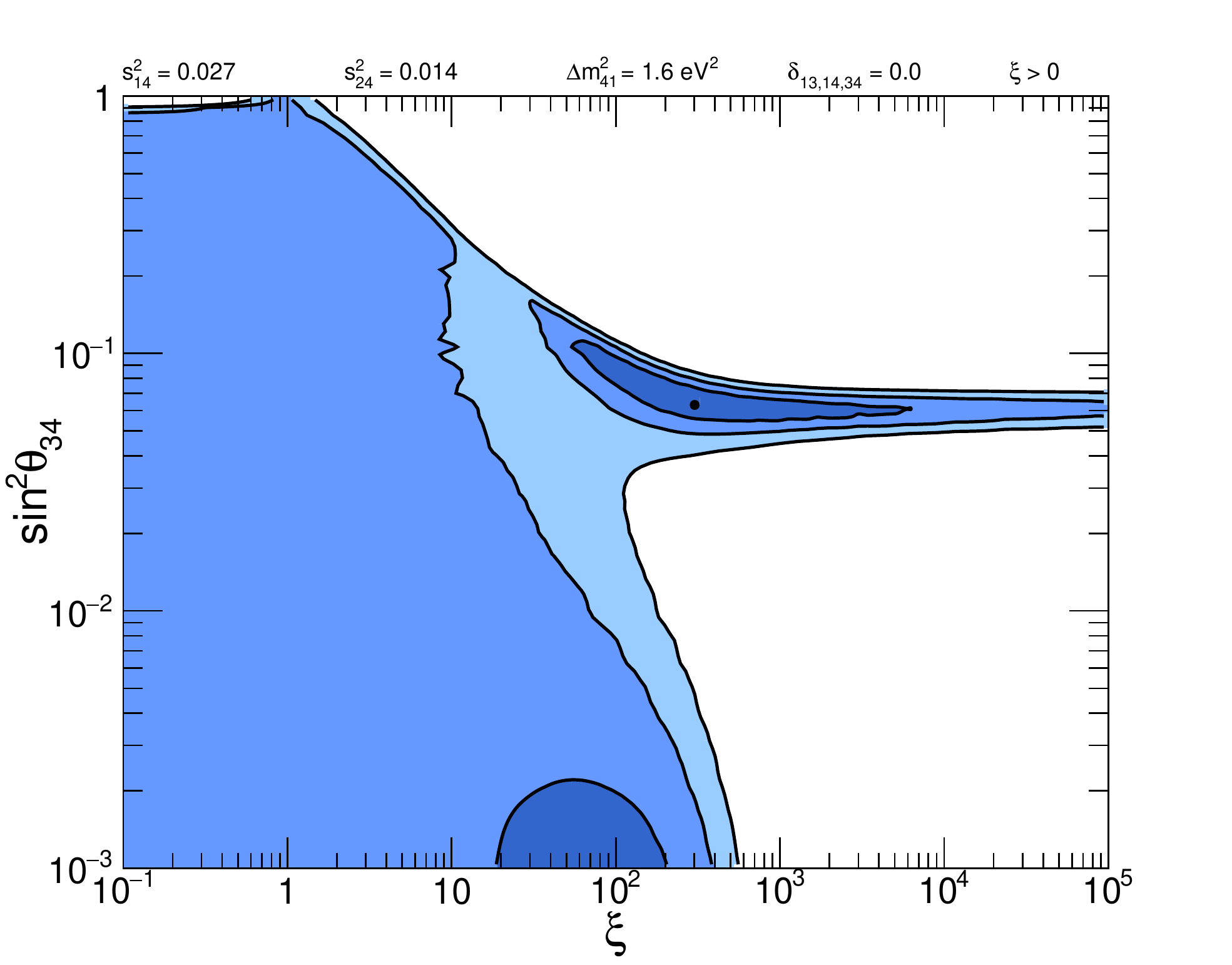}~~\includegraphics[width=8cm]{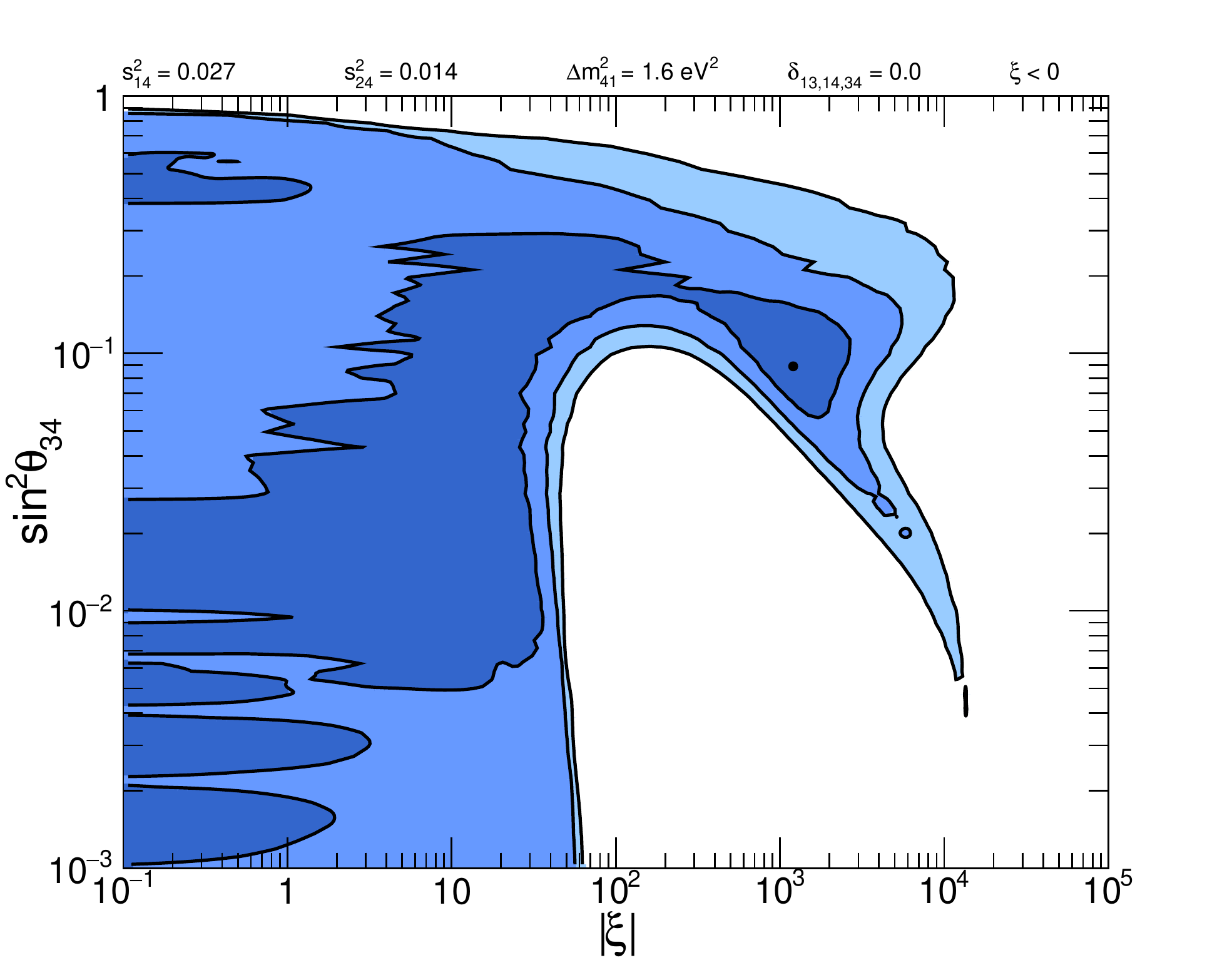}
\caption{Allowed $s_{34}, \xi$ contour regions with $\xi>0$ (left) and $\xi<0$ (right). Here $s_{14}=0.027$, $s_{24}=0.014$, and $\delta_{13,14,34}=0$. The same confidence
levels reported in Fig. \ref{solarFit} are used.
}
\label{34GX}
\end{center}
\end{figure}

Figure~\ref{34GX} shows a scan of $s^2_{34}, \xi$ with the other parameters fixed to $s^2_{14}=0.027, s^2_{24}=0.014$ and $\delta_{13,14,34}=0$. The qualitative behavior is similar to Fig.~\ref{idea}, except that $s^2_{34}$ does not affect the overall scaling of $P_{ee}$ and larger angles are allowed.

Lastly, it is instructive to show how the Dark MSW effect shows up in the survival probability $P_{ee}$. For a given neutrino flux component $i=$ pp, O, N, $^7$Be, hep, $^8$B, we denote by $\phi_i(E)$ and $\rho_i(r)$ the flux per unit energy and the average production point, respectively. By definition $\int dr\, \rho_i(r)=1$, whereas $\int dE~\phi_i=\Phi_i$ is the total flux. Terrestrial detectors are sensitive to $\sum_i\int dr~\phi_i\rho_iP_{ee}(r,E)$. However, given the huge difference in magnitude among the total fluxes $\Phi_i$, plotting this quantity is not convenient for our purposes. To better illustrate the distinctive features of our framework it is instead more useful to consider the averaged (day) survival probability:
\ba\label{averaged}
\overline{P}_{ee}(E)=\int dr\, P_{ee,{\rm day}}(r,E)\frac{\sum_i\phi_i(E)\rho_i(r)}{\sum_i\phi_i(E)}.
\ea
\begin{figure}[t]
\begin{center}
\includegraphics[width=8cm]{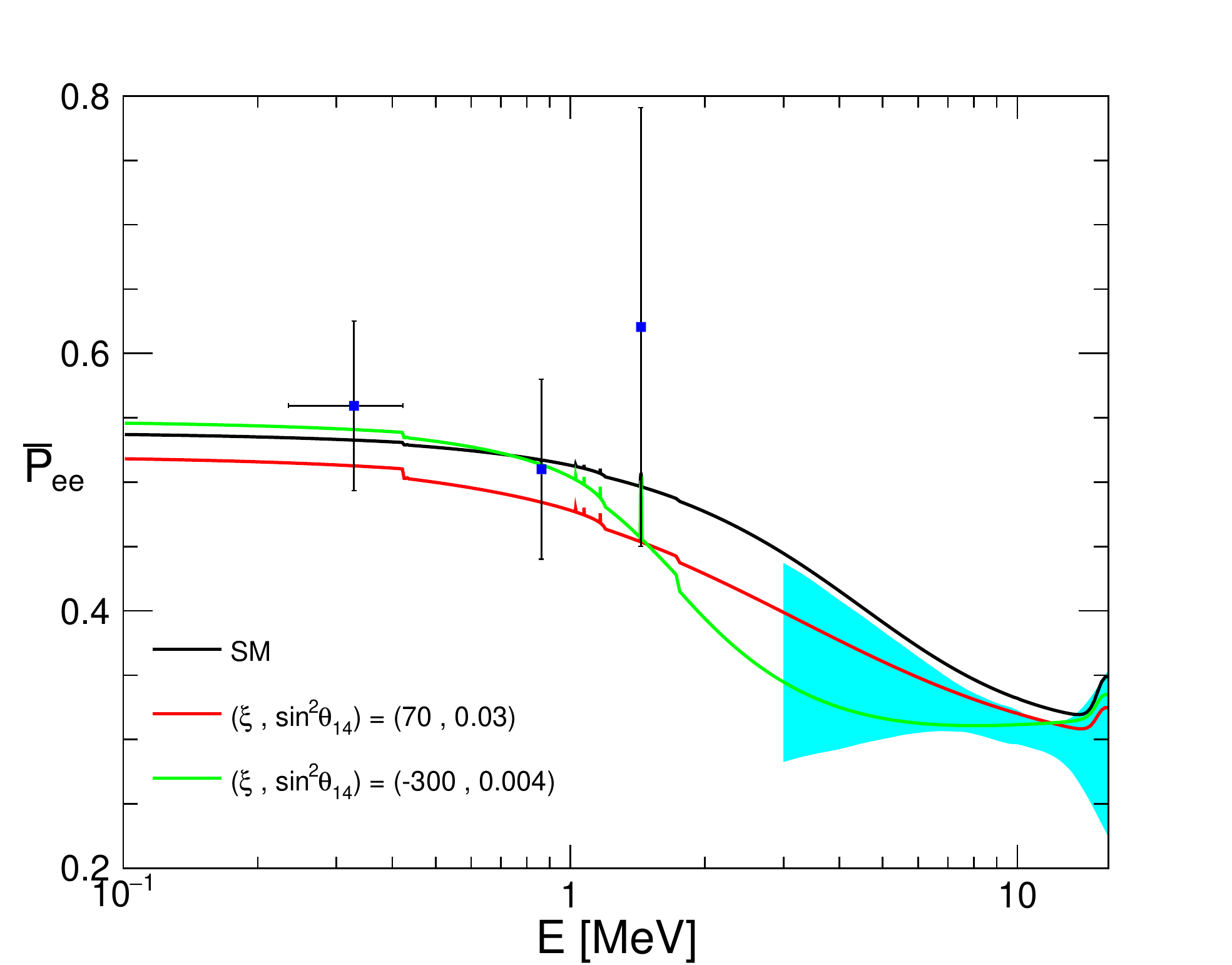}~~\includegraphics[width=8cm]{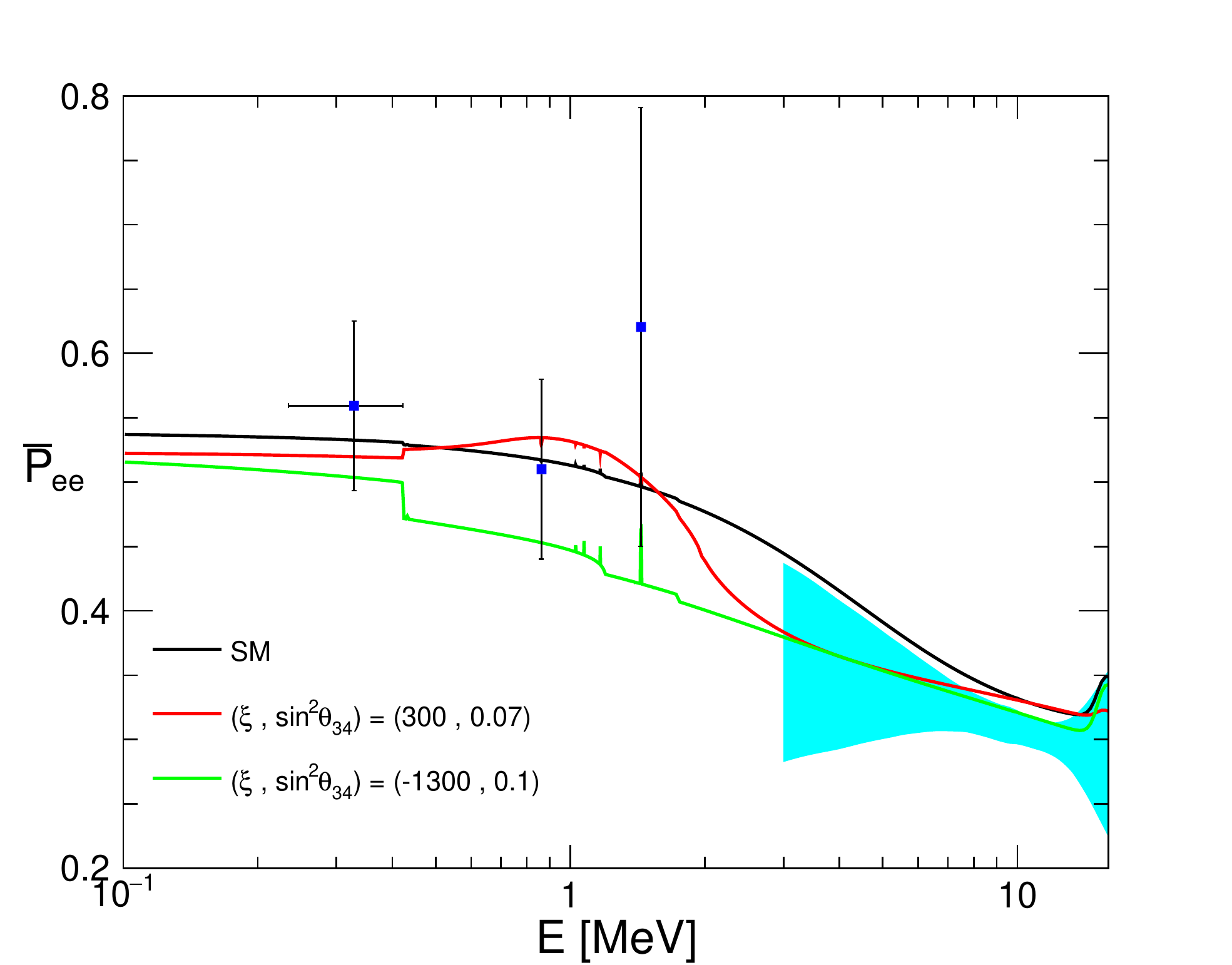}
\caption{$\overline{P}_{ee}(E)$ (see Eq.(\ref{averaged})) for a few benchmark points with $\xi>0$ (red lines) and $\xi<0$ (green lines). The numerical values of $s_{14,24}$, when not explicitly shown in the plot, as well as $\Delta m_{41}^2$ are taken from the best fit values of \cite{Gariazzo:2015rra} ($s_{34}$ and all phases are set to zero) whereas the standard parameters are $s_{12}^2=0.3$, $\Delta m_{21}^2=7.5\times10^{-5}$ eV$^2$. The black curve is the Standard Model prediction. The data points refer (from left to right) to pp (digitalized from \cite{Aharmim:2011vm}), $^7$Be, pep (from Borexino~\cite{Bellini:2013lnn}), and the SNO $^8$B neutrinos band.
}\label{PeeFig}
\end{center}
\end{figure}
We show it for a few benchmark points in Figure \ref{PeeFig}. For comparison the Standard Model prediction is overlaid in black. $\overline{P}_{ee}(E)$ manifests a set of discontinuities at the transition between neutrino species. The larger the new physics effects, the bigger the discontinuous jumps. This is the main signature of Dark-MSW. As expected, the departure from the black line is more evident in the $^8$B, $^7$Be, CNO regions, and far less so for pep and pp neutrinos. In particular, note the sharp effect at pep (third data point in the figure), where our red and green curves spike towards the black line. Similarly, the irregularities observed in the N region (between the second and third data point in Fig.~\ref{PeeFig}) arise because these neutrinos are produced partly centrally, where our effect is important, and partly at $>15\% R_\odot$, where Dark MSW is negligible and $P_{ee}$ approaches the black curve (see black dashed curve in Fig.~\ref{densities}).

Note also a bump in the red curve of Eq.~\ref{PeeFig}, right. Its origin is a jump of reduced 1-2 mixing where $P_{ee}\sim \cos^2\theta_{12}$ that was also noted in a similar context in~\cite{Kopp:2014fha}. This can be understood from our analytic expression of Section Eq.~\ref{sec:analyt} as a result of a cancellation leading to $|\Delta \sin2\theta_{12}+V_y|\ll|\Delta \cos2\theta_{12}-V_x|$. The effect gets smeared out when $P_{ee}$ is convolved with the production point in Eq.~(\ref{averaged}).

\section{Conclusions}
\label{sec:conc}

New phenomena in neutrino oscillations may be introduced via a mixing with sterile neutrinos carrying exotic couplings to Dark Matter. These scenarios, schematically depicted in Fig.~\ref{fig:scheme}, are motivated both phenomenologically and theoretically. On the one hand, anomalies in short baseline experiments motivate the existence of eV scale sterile neutrinos with $\theta_s\sim0.1$ mixing with active ones. On the other hand, theoretical considerations suggest that such steriles must possess exotic interactions in order to avoid conflicts with standard Cosmology. The latter may naturally involve Dark Matter. 

We pointed out that a novel, characteristic signature of the resulting picture is a Dark MSW effect: an exotic matter potential that neutrinos can feel when traveling through an asymmetric DM medium. We focused on asymmetric DM scenarios with $m_X\sim5-15$ GeV and argued that Dark MSW is controlled by the product of three unknown parameters: $\sin^2\theta_s~G_Xn_X$, with $\sin\theta_s$ the active-sterile mixing, $n_X$ the DM charge density, and $G_X$ a dimensionful quantity set by the mass scale of the new physics, in analogy to $G_F$ in the Standard Model. Crucially, for sufficiently light exotic sectors, $G_X$ can be extremely large and compensate the possibly small density. This implies that Dark MSW might represent a key signature in the limit of small dimensionless exotic couplings and light mediators, when all other probes of neutrino-DM interactions become inefficient, see Fig.~\ref{fig:dMSW}.

Dark MSW can affect oscillations of highly energetic neutrinos traveling through the cosmos. However, the flavor content of such neutrinos is not well understood and no constraint can be derived in that case. Yet, we pointed out that a more promising signature of Dark MSW is possible if a fraction of DM has accumulated in the Sun. In that case a phenomena dubbed {\emph{Solar Dark MSW}} can take place.

DM capture by the Sun --- which requires a non-vanishing nucleon-DM interaction $\sigma_{nX}$ --- has been investigated by many groups, for example as a possible solution to the so-called ``solar abundance problem" or simply as a source of indirect signatures of DM. Solar Dark MSW, however, does not necessitate the large DM density invoked to solve the ``solar abundance problem" and, at the same time, is present only in asymmetric DM models, where indirect signatures of DM annihilation are suppressed. In this sense, Solar Dark MSW is a new physics effect {\emph{orthogonal}} to the ones discussed in the previous literature.

The main feature characterizing our scenario is the ability to modify solar neutrinos quite substantially via the Solar Dark MSW, and simultaneously hide --- or behave as an ordinary sterile neutrino framework --- in all other respects. Because by construction dark matter gets localized in the Sun's core, it is mostly the physics of $^8$B, $^7$Be, and CNO neutrinos that is modified, whereas pp and pep neutrinos remain largely unaffected even for sizable exotic couplings $G_X$, see Fig.~\ref{PeeFig}. Interestingly, assuming exotic mass and mixing angles at the values suggested by short baseline anomalies, we find that a solar mass difference $\Delta m_{21}^2$ in the KamLAND range is naturally preferred when $G_X>0$ whereas, similarly to NSI, a Dark-LMA solution is mildly favored for $G_X<0$. In general, large departures from the Standard Model in the CNO spectra may be interpreted as a smoking gun of the present scenario. Currently, solar neutrino data allow exotic potentials $10$ times larger than what is expected in the Standard Model,
\ba
s_{i4}^2|\xi|={\cal O}(10),
\ea
where $\xi\sim G_Xn_X/G_Fn_e$, but even larger $s_{i4}^2|\xi|$ are possible when several exotic angles are turned on (in which case resonances involving $\Delta m_{31}^2$ may occur). Such large deviations are allowed precisely because Solar Dark MSW impacts the transition region between the vacuum- and matter-dominated regimes of $P_{ee}$, where Kamland and solar data manifest a mild tension within the standard 3-neutrino framework.

Our main message here is that solar data are capable of probing a large portion of the parameter space of the models of Fig.~\ref{fig:scheme} when all other bounds are completely ineffective, see Fig.~\ref{fig:dMSW}. In this paper we focused on scenarios with $\sin\theta_s\sim0.1$ and $\Delta m_{41}^2\sim1$ eV$^2$, as motivated by the short baseline anomalies, but our conclusions are quite general. An analysis of the entire parameter space is beyond the scope of our paper and is left for future work.

We conclude stressing that while the sterile neutrinos play a crucial role in our models, they do not represent a general signature. The steriles are simply messengers between the visible and dark worlds. In some scenarios they may be too heavy or too weakly-coupled to the Standard Model to be directly visible, and in that case Dark MSW may nevertheless be relevant. Indeed, an examination of Eq.~(\ref{hamilton}) reveals that Dark MSW admits a {\emph{decoupling limit}}
\ba\label{decoupling}
\sin\theta_s\to0~~~~~~~~~~~~~~~\sin^2\theta_s~\xi={\rm finite},
\ea
in which the sterile neutrinos effectively disappear while leaving an exotic potential for the active neutrinos as their only imprint. To see this more explicitly, observe that as $y_a\langle H\rangle/y_s\langle\phi\rangle\to0$ with $\sin^2\theta_s~\xi$ held fixed neutrino oscillations are described by the standard 3 by 3 Hamiltonian (see the upper left 3 by 3 block in Eq.~(\ref{hamilton}) with vacuum mixing matrix $U=R_{23}U_{13}R_{12}$) plus an additional nonstandard potential term given by:
\ba\label{Hnew}
H^{\rm new}_{ij}=V_s(U_{34}R_{24}U_{14})^*_{4i}(U_{34}R_{24}U_{14})_{4j}\to G_Xn_X\theta_{i4}\theta_{j4}.
\ea 
(After $\to$ we set $\delta_{14,34}=0$ for simplicity. Also, note that off-diagonal matter potentials require at least two non-vanishing angles $\theta_{i4}$, a feature that was also emphasized in Section~\ref{sec:analyt}.) The potential in Eq.~(\ref{Hnew}) has the same appearance of NSI, provided we identify $G_Xn_X\theta_{i4}\theta_{j4}=V_{\rm CC}\varepsilon_{ij}$. 

Our framework generalizes the more conventional neutrino-NSI in many respects. First, light sterile neutrinos may provide additional crucial signatures. Second, in our framework the exotic potential is controlled by other forms of matter. This means Dark MSW is only relevant in neutrino propagation (and not at production/detection), and furthermore we are effectively promoting $\varepsilon_{ij}$ to position-dependent parameters. Third, the range the effective $\varepsilon_{ij}$ are allowed to vary in is much larger here. In fact, as opposed to our framework, traditional NSI cannot allow the large couplings displayed in Figs.~\ref{idea} and \ref{34GX} --- whose effect becomes particularly manifest in the ``trunk-'' and ``horn-like'' regions of our plots --- because constrained by a myriad of other terrestrial experiments that force them to be small corrections of the Fermi interactions.

\section*{Acknowledgments}

L.V.~acknowledges the MIUR-FIRB grant \sloppy\mbox{RBFR12H1MW} and the ERC Advanced 
Grant no.267985 ({\emph{DaMeSyFla}}). F.C.~acknowledges the research grant 2012CPPYP7 
under the program PRIN 2012 funded by the Italian Ministry of Education, University and 
Research (MIUR). F.C.~also acknowledges support from the
NSF Grant PHY-1404311 to J.F. Beacom.

\appendix

\section{A completion of (\ref{Lagrangian}) with accidentally massless $\nu_s$}
\label{sec:anomalyfree}

Here we present a simple UV complete theory whose low energy phenomenology contains a massless Dirac fermion $\nu_s$ charged under a gauge $U(1)$, as in (\ref{Lagrangian}). 

Consider an exotic gauge group $SO(2N+1)\times SU(2)$ and introduce the anomaly-free field content of the following table:

\begin{table}[h]
\begin{center}
\begin{tabular}{c|ccc} 
\rule{0pt}{1.2em}%
fields & $SO(2N+1)$ &  $SU(2)$ & Lorentz   \\
\hline
$\nu_s$ & ${\bf 1}$ & ${\bf 2}$ & Weyl \\
$\Psi$ & ${\bf 2N+1}$ & ${\bf 2}$ & Weyl \\
$\phi$ & ${\bf 1}$ & ${\bf 2}$ & Scalar 
\end{tabular}
\end{center}
\end{table}
At the renormalizable level, no mass terms for $\nu_s,\Psi$ nor Yukawa couplings can be written down consistently with the gauge symmetries. This implies the existence of accidental global symmetries with important physical consequences, as described below.

The $SO(2N+1)$ dynamics is assumed to confine producing a condensate $\langle\Psi\Psi\rangle$. The latter breaks $SU(2)\to SO(2)\sim U(1)$ with the associated composite Nambu-Goldstone modes becoming the longitudinal components of the broken gauge generators. Importantly, the strong dynamics enjoys an accidental $G$-parity acting on $\Psi$ and the $SO(2N+1)$ gauge bosons. As a consequence, some of the composites are stable. To avoid over-closing the universe we assume the phase transition occurred before inflation, such that any abundance of the heavy composites was depleted at reheating. Furthermore, the massive gauge boson is assumed to contribute negligibly to dark matter, for simplicity. If the $SU(2)$ coupling is sizable, this follows because the vector is heavy and therefore its abundance is washout after inflation; if the $SU(2)$ coupling is small the vector was not populated in the early Universe because too weakly-coupled. In this framework the field denoted by $X$ in (\ref{Lagrangian}) is part of an additional Dark Sector with vector-like $SU(2)$ charges and a primordial dark asymmetry.

Finally, under the unbroken $U(1)$ gauge theory $\nu_s$ describes a Dirac fermion. However, because of an accidental (anomalous) axial symmetry the steriles $\nu_s$ remain massless, as in (\ref{Lagrangian}). The singlet fermion of lowest dimension is $\phi\nu_s$, as assumed in (\ref{Lagrangian}).

\section{An alternative parametrization for $U$, with $s_{34}=0$}
\label{sec:alt}

The parametrization of the vacuum mixing matrix used in the main text is convenient when comparing to other work and because the expression for $P_{ee}$ is particularly simple. However, agreement with the numerical solution is not excellent. The reason is that the off-diagonal terms in $H_{i3}$ are proportional to $V_s$. Hence, the approximation that $U'\approx VU_{12m}$ diagonalizes the full Hamiltonian breaks down when $s_{i4}^2V_sE/\Delta m_{31}^2$ approaches one. Large portions of the parameter space with sizable $\xi$ are therefore not well described by this approximation.

Consider instead $U=R_{23}U_{13}U_{14}R_{24}U_{34}R_{12}$, where $s_{34}$ and all phases are taken to vanish for definiteness. In this case one can show that $H'_{i3}\propto V_{\rm CC}$ whereas $H'_{i4}\propto V_{s}$. As a result, the error encountered in taking $U'\approx VU_{12m}$ becomes of order $V_{\rm CC}E/\Delta m_{31}^2$, which is significantly smaller than in the standard parametrization.

When $V_{\rm CC}E/\Delta m_{31}^2\ll1$ the problem can be described by the effective hamiltonian (\ref{eff}), where now
\ba
V_x(r,E)&=&\frac{1}{2}[V_{\rm CC}c_{13}^2(c_{14}^2-s^2_{14}s_{24}^2)+V_s(s^2_{14}-c^2_{14}s_{24}^2)]\\\no
V_y(r,E)&=&(V_s-V_{\rm CC}c_{13}^2)c_{14}s_{14}s_{24}.
\ea
The survival probability has a more involved expression: 
\ba\label{improved}
P_{ee}&=&s_{13}^4 + 
 c_{13}^4 c_{24}^4 s_{14}^4 \\\
 &+& 
 c_{13}^4 (c_{14}s_{12} - 
    c_{12}s_{14} s_{24})^2
(c_{14} s_{m} - 
    c_{m} s_{14} s_{24})^2 \\\no
    &+& 
 c_{13}^4 (c_{12} c_{14} + 
    s_{12} s_{14} s_{24})^2
(c_{m} c_{14} + 
    s_{m} s_{14} s_{24})^2\\\no
    &+&{\cal O}(V_{\rm CC}E/\Delta m_{31}^2).
\ea
with $\theta_{m}$ the same as given in Eq.~(\ref{analtheta}). The improved precision of Eq.~(\ref{improved}) allowed us to perform an accurate check of our numerical simulation.

\section{Details on the numerical analysis}
\label{sec:chi}

In order to construct our $\chi^2$, we first calculate $P_{ee,{\rm day}}$ and $P_{ee,{\rm night}}$ as explained in the text and then convolve these quantities with the relevant production point distribution $\rho_i(r)$:
\ba
\langle P_{ee}^i({\bf p},E)\rangle\equiv\int dr\rho_i(r) P_{ee}({\bf p},r,E),
\ea
where $i=^8$B for $\chi^2_{\rm SNO}({\bf{p}})$ and $i=^7$Be, pep for the last two terms in Eq.~(\ref{chi}), respectively. In particular, for the $^7$Be (pep) case the energy is fixed to 862 (1440) keV. The last two terms of Eq.~(\ref{chi2}) are then defined as
\ba\no
\chi^2_{\rm Be7}({\bf{p}})&=&\left(\frac{\langle P^{^7\rm{Be}}_{ee}({\bf p},862~{\rm keV})\rangle-0.51}{0.07}\right)^2\\\no
\chi^2_{\rm pep}({\bf{p}})&=&\left(\frac{\langle P^{\rm{pep}}_{ee}({\bf p},1440~{\rm keV})\rangle-0.62}{0.17}\right)^2,
\ea
where the experimental values of the survival probability and its errors are taken from Borexino~\cite{Bellini:2013lnn}.

The calculation of $\chi^2_{\rm SNO}(\bf p)$ is more involved. The SNO collaboration provides a $6\times6$ correlation matrix (we indicate with $\Sigma$ the correspondent covariance matrix) regarding the total $^8$B flux $a_0$, the three coefficients $a_1,a_2,a_3$ of the quadratic polynomial obtained by analyzing their data collected during the day with a polynomial expression of $P_{ee,{\rm day}}$, and the two coefficients $a_4,a_5$ related to the linear fit of the day-night asymmetry. The best fit values of the same parameters are also available. In order to use the information given by the collaboration, we need to evaluate $a_i(\bf{p})$ for the point of the parameter space $\bf p$ under consideration. First, we note that the estimation of $P_{ee}$ in SNO occurs through the ratio between charged current and neutral current events, where the last ones include only active neutrino flavors. We then define our theoretical prediction of the observed survival probability in SNO as
\ba
\langle P_{ee}^{\rm SNO}({\bf p},E)\rangle=\frac{\langle P^{^8{\rm B}}_{ee}({\bf p},E)\rangle}{1-\langle P^{^8{\rm B}}_{es}({\bf p},E)\rangle}\ .
\label{Pee_SNO}
\ea
The same form applies to both the day and night probabilities. A quadratic fit of $\langle P_{ee,{\rm day}}^{\rm SNO}({\bf p},E)\rangle$ allows to calculate $a_i(\bf p)$ ($i=1, 2, 3$), whereas $a_i(\bf p)$ ($i=4, 5$) are obtained through a linear fit of the day-night asymmetry $A({\bf p},E)$, defined as
\ba
A({\bf p},E) =2 \frac{\langle P^{^8{\rm B}}_{ee,{\rm night}}({\bf p},E)\rangle-\langle P^{^8{\rm B}}_{ee,{\rm day}}({\bf p},E)\rangle}{\langle P^{^8{\rm B}}_{ee,{\rm night}}({\bf p},E)\rangle+\langle P^{^8{\rm B}}_{ee,{\rm day}}({\bf p},E)\rangle}.
\ea
As far as $a_0(\bf p)$ is concerned, if we denote by $\Phi$ the predicted value of total $^8$B neutrino flux from the Standard Solar Model, then $a_0({\bf p})=\Phi(1-\langle P^{^8{\rm B}}_{es} \rangle)$, where the average here is performed over the production point, the energy and both the day and night cases. Finally, we can write
\ba
\chi^2 _{\rm SNO}({\bf p})={\rm min}_{\Phi}\left\{\sum_{i,j=0}^5\left[a_i({\bf p})-a_i^{\rm SNO}\right]\Sigma_{ij}^{-1}\left[a_j({\bf p})-a_j^{\rm SNO}\right]+\chi^2_{\rm SSM}(\Phi)\right\}\ ,
\label{chi2_SNO}
\ea
where we marginalize over $\Phi$, considering a range of $\pm5\sigma$ according to the error reported by the collaboration, and $\chi^2_{\rm SSM}(\Phi)$ represents a quadratic penalty taking into account the constraints from the Standard Solar Model. Here we adopt the model BPS09(AGSS09) of \cite{Serenelli:2009yc}.

The above procedure reproduces reasonably well the Standard Model  fit (see dashed contours in Fig.~\ref{solarFit}), which we take as a confirmation of the accuracy of our approach. We further verified that the polynomial parametrization is very accurate in all scanned regions, even when resonances involving $\nu_3$ are present.

\bibliographystyle{JHEP}

\bibliography{nu}

\end{document}